\documentclass[pra,nofootinbib,superscriptaddress,showpacs,showkeys]{revtex4}

\usepackage{graphicx,epsfig}
\usepackage{amsfonts,amsmath,amssymb,amsthm,amscd}

\begin{document}

\title{Damping of electromagnetic waves due to electron-positron pair
production}

\author{S. S. Bulanov}
\email{bulanov@heron.itep.ru} \affiliation{Institute of
Theoretical and Experimental Physics, 117218 Moscow, Russia}

\author{A. M. Fedotov}
\email{fedotov@cea.ru} \affiliation{Moscow State Engineering
Physics Institute, 115409 Moscow, Russia}

\author{F. Pegoraro}
\email{pegoraro@df.unipi.it} \affiliation{Department of Physics,
University of Pisa and INFM, Pisa, Italy}

\begin{abstract}
The problem of the backreaction during the process of
electron-positron pair production by  a circularly polarized
electromagnetic wave propagating in a plasma is investigated. A
model based on the relativistic Boltzmann-Vlasov equation with a
source term corresponding to the Schwinger formula for the pair
creation rate is used. The damping of the wave, the nonlinear
up-shift of its frequency due to the plasma density increase and
the effect of the damping on the wave polarization and on  the
background plasma acceleration  are investigated as a function of
the wave amplitude.
\end{abstract}

\maketitle

\section{Introduction}

The production  of  electron-positron pairs under the action of
electromagnetic field (Schwinger effect) attracts great attention
because it is a nonlinear effect that lies beyond the limits of
perturbation theory. The study of this effect can shed light on
the nonlinear properties of the quantum electrodynamics (QED)
vacuum.

This effect was first predicted for the case of  a constant
electric field more than 60 years ago \cite{Saut} (see also
\cite{el-pos,Schwinger}). It is well known that a plane
electromagnetic wave cannot produce electron-positron pairs
because  both its electromagnetic invariants, $(E^2 - B^2)/2$ and
$\mathbf{E}\cdot \mathbf{B}$, are equal to zero. For this reason
this effect was first considered in the case of a constant
electric field, in which case the first invariant $(E^2 - B^2)/2$
does not vanish. Later, this analysis was extended to the case of
a spatially homogeneous time-varying electric field
\cite{B-I,Popov71,Popov73, PopovMarinov,NN}, but  these results
were long believed to be of academic interest only, because the
power of the laser systems available at that time was far below
the limit  for   pair production  to become experimentally
observable\cite {Popov73, Bunkin, Troup}. However the recent
development of laser technology has resulted in the increase of
the power of optical and infrared lasers  by many orders of
magnitude \cite{Mourou}.  Presently, lasers systems are available
that can deliver pulses  with  intensities  of the order of
$10^{22}$ W/cm$^2$ in the focal spot. Such intensities are still
much smaller than the characteristic intensity for pair production
$I_{Sch}=4.6 \times 10^{29}$ W/cm$^2$, which corresponds, for a
laser pulse with wavelength $\approx 1 \mu m$,  to an electric
field equal  to the critical Schwinger  field $E_{Sch}=1.32\times
10^{16}$ V/cm.  Nevertheless, there are projects that aim to reach
intensities as high as $10^{26} - 10^{28}$ W/cm$^2$ already in the
coming decade. In addition, several methods for reaching the
critical intensity with presently available systems have been
proposed recently. One of these schemes was demonstrated in the
experiments at SLAC where $10^{18}$ W/cm$^2$ laser photons,
back-scattered by a $46.6$ GeV electron beam, interacted with the
laser pulse and several electron-positron pairs were detected
\cite{Burke}. Another scheme for reaching critical intensities was
suggested in Ref.\cite{LightIntensification}, where the
interaction  of the laser pulse with electron density modulations
in a plasma, produced  by  a counterpropagating breaking wake
plasma wave, results in the frequency up-shift and pulse focusing.
In this scheme intensities of the order of the critical density
can be obtained using   $10^{18}$ W/cm$^2$ laser pulses. Hence,  a
more detailed study of the Schwinger effect  in time-varying
electromagnetic fields  and of all the processes that accompany it
has become an urgent physical problem from an experimental point
of view also.

The process of electron-positron pair production by
electromagnetic fields that are solutions of the Maxwell equations
in a plasma and in vacuum was studied recently  in Refs.
\cite{ssbul,BNMP}. In Ref. \cite{BNMP} it was shown that already
for intensities of the laser pulse  smaller than the critical
intensity, the energy loss due to pair production is of the same
order of the energy storied in the pulse. Therefore it is no
longer possible to consider the electromagnetic field in the pulse
as an external field and  the energy loss by the electromagnetic
field due to pair production and particle acceleration must be
taken into account.

The problem of the backreaction of the produced particles on the
background field was discussed extensively in a number of papers
on the particle formation process in  high energy hadronic
interactions \cite{Matsui,Matsui1,Kluger,Schmidt} as well as under
the action of electric fields \cite{Ruffini,Dolby}. In the former
case this process can be viewed as the quantum tunneling of
quark-antiquark and gluon pairs in the presence of the background
color-electric field of quantum chromodynamics (QCD). Such a color
field is formed between two receding nuclei which are color
charged by the exchange of soft gluons at the time of collision.
This leads to the formation of a very strong color-electric field
and, hence, to more copious pair production. It was understood
that, in solving a dynamical problem with a strong initial
electric field, the effect of the produced particles on the
electric field (the back reaction) should be taken into
consideration. The quark-gluon plasma emerging through tunneling
in nucleus-nucleus collisions will change the color-electric field
due to the appearance of conduction and polarization currents. The
first current  is due to the particle motion in the field while
the latter  arises from the process of pair creation.  A kinetic
equation coupled to Maxwell equations was used to solve this
problem. The pair production process was considered as if occurring in QED, and
the color-electric field was assumed to be Abelian, spatially
homogeneous and time-dependent. However a spatially homogeneous
time dependent electric field is not a solution of Maxwell
equations in vacuum. We  also note that in Refs.\cite{Matsui,
Matsui1, Kluger, Schmidt, Dolby, Ruffini} special attention was
paid to the properties of the emerging plasma, while the
properties of the  background field were not  studied in detail.

In the present paper, we consider the process of electron-positron
pair production in a cold collisionless plasma, under the action
of an electromagnetic field which is an actual  solution of the
Maxwell equations, as well as the backreaction of the produced
pairs on the background field. In doing so we use the
Boltzmann-Vlasov equation, with a source term obtained from the
pair production rate \cite{Matsui,Matsui1,Kluger}. In order to
elucidate the role of the magnetic field component on the
electron-positron pair production, we consider a planar,
circularly polarized, electromagnetic wave propagating in an
underdense collisionless plasma (for the sake of simplicity we
consider an electron-positron plasma). In the case of a plane wave
in  a plasma the first field invariant $(E^2 - B^2)/2$ is not
equal to zero  due to the different dispersion equation with
respect to that in vacuum.  Therefore, in  a plasma,
electron-positron pairs can be produced by  a plane
electromagnetic wave, as  was shown in Ref.\cite{ssbul}. In this
case a Lorentz transformation to the reference frame moving with
the group velocity $v_{g}$ of the wave transforms the
electromagnetic field into a purely electric field, that rotates
with constant frequency, and with no associated magnetic field.
Although this transformation reduces the problem under
consideration to the situation where the pairs are produced by a
time-varying electric field, the effects of the wave magnetic
field  are incorporated rigorously into our model. We notice that
a similar approach was used  earlier in Ref.\cite {avetissian}.
However in this latter paper a linearly polarized electromagnetic
wave was considered and an approximation was adopted that is only
valid in the limit of a small amplitude electromagnetic field.

In the present paper we consider the effect on the background wave
in the plasma caused by  the  pairs produced by the wave  through
their polarization and conduction currents. In particular, by
considering the interaction between the wave and the plasma as an
initial value problem in the moving frame,we study the
evolution of the wave electromagnetic field. Due to the nonlinear
properties of the equations governing the field evolution in time,
we find a strong nonlinear dependence of wave field properties on
the wave initial amplitude. We find a nonlinear up-shift of
the wave frequency, a change of its polarization state and damping
of its amplitude. In order to exemplify these effects, we consider
two limiting cases of the electric field evolution. In the first
limit the amplitude of the initial electric field is assumed to be
so large that most of the electron-positron pairs are produced
instantaneously. This leads to an  instantaneous change of the
electric field amplitude and frequency. In the second limit  we
consider a regime where the pair production rate is relatively
small, so that all parameters of the wave change slowly. In this
case we see the wave amplitude damp with time.

This  paper is organized as follows. In Sec.\ref{RS} we review
some well known  properties of  a strong electromagnetic wave in a
plasma. In Sec.\ref{KD} we study the equations governing the
process of pair production and  the evolution of the
electromagnetic wave. In Secs.\ref{L1},\ref{L2} we consider two
limiting cases: fast changing field and slow changing field. The
main results and conclusions are presented in Sec.\ref{CC}.

\section{Relativistically strong electromagnetic wave in a plasma}\label{RS}

First we recover the properties of a relativistically strong
electromagnetic wave  propagating in an underdense collisionless
electron-positron plasma that are needed in order to determine the
form  of the time varying electric field.  This  discussion is
based on the results obtained by Akhiezer and Polovin in Ref.
\cite{AP}.  We consider a circularly polarized electromagnetic
wave, propagating in an electron-positron plasma. In the following
we use the $c=1$ and $\hbar =1$ convention.

An electromagnetic wave propagating in a plasma has a group
velocity smaller than the speed of light in vacuum. Thus it is
possible to make a transformation to the reference frame moving
with the wave group velocity  $v_{g}$. In this frame the magnetic
field of the wave vanishes and its time-varying electric field is
spatially homogeneous and is thus governed by the equation
\begin{equation}
\frac{d{}\mathbf{E}}{dt}=-4\pi \mathbf{j}=-4\pi
\sum\limits_{\alpha =+,-}\frac{e_{\alpha }}{(2\pi )^{3}}\int
\mathbf{v}_\alpha f_{\alpha }(\mathbf{p},t)d^{3}p.
\label{el-field}
\end{equation}
where $\mathbf{v}{}=\mathbf{p}{}/(m^{2}+p^{2})^{1/2}$,
$p=|\mathbf{p}|=(p_x^2+p_y^2+p_z^2)^{1/2}$, $f_{\alpha
}(\mathbf{p},t)$ is the positron (electron) distribution function,
normalized such that $\int f_{\alpha }(\mathbf{p},t)d^{3}~p/(2\pi
)^{3}=n_{\alpha }$ gives the number $n_{\alpha }$  of electrons or
positrons per unit volume, and $e_{\alpha }$ is their electric
charge   with $\alpha =+$ for the positrons  and $\alpha =-$ for
the electrons. We consider the case of an electrically
quasineutral plasma, where the density of positrons is equal to
the density of electrons, $n_{+}=n_{-}=n_{0}$, with $n_{0}$ the
density in the moving frame. If we assume that the plasma is cold,
i. e., that the particle distribution function of the species
$\alpha $ can be written as $f_{\alpha}(\mathbf{p},t)=n_{\alpha
}{(2\pi )^{3}}\delta (\mathbf{p}-\mathbf{p}_{\alpha }(t))$, the
system of equations for the electric field evolution in the moving
frame reduces to
\begin{eqnarray}
\frac{d{}\mathbf{E}}{dt} &=&-4\pi n_{0}\sum\limits_{\alpha
=+,-}{e_{\alpha }} \mathbf{v}_{\alpha },\label{JJ} \\
\frac{d{}\mathbf{p}_{\alpha }}{dt} &=&e_{\alpha
}\mathbf{E}{},\quad {\rm with} \quad \mathbf{v}{}_{\alpha }
=\frac{\mathbf{p}{}_{\alpha }}{(m^{2}+p_{\alpha }^{2})^{1/2}}.
\end{eqnarray}
We assume symmetric initial conditions, so that the positron and
the electron momenta ${p}_{\pm,\bot} $ perpendicular to the
direction of propagation of the laser pulse  have opposite signs
and equal absolute value, $\mathbf{p}_{+,\bot}=
-\mathbf{p}_{-,\bot}=\mathbf{p}_\bot$, while they have equal
parallel momentum ${p}_{\Vert} = {p}_{\Vert,0}$. Here the non zero
value $p_{\Vert ,0}=-mv_{g}\gamma _{g}$ of the parallel momentum
is due to the Lorentz transformation from the laboratory to the
moving  frame and $\gamma_{g}=(1-v_{g}^{2}/c^{2})^{-1/2}$.  Then,
we find for the particle momentum $\mathbf{p}$ the equation
\begin{equation}
\frac{d^{2}\mathbf{p}_\bot}{dt^{2}}=-\frac{8\pi
e^{2}n_{0}\mathbf{p}_\bot}{(m^{2}+p^{2})^{1/2}}
\end{equation}
The solution of this equation is
\begin{equation}
{}\mathbf{p}_\bot=-P(\mathbf{e}_{x}\cos \Omega
t+\mathbf{e}_{y}\sin \Omega t), \quad {\rm and }\quad
\mathbf{A}=A_{0}(\mathbf{e}_{x}\cos \Omega t+\mathbf{e}_{y}\sin
\Omega t),
\end{equation}
where $\mathbf{A}$ is the  vector-potential,  $A_{0}=P/e$ and
\begin{equation}
\Omega =\left[ \frac{8\pi e^{2}n_{0}}{\left( m^{2}+P^{2} + p_{\Vert
,0}^2 \right)
^{1/2}}\right] ^{1/2}
\end{equation}
is the Langmuir  frequency which enters the dispersion equation of
an electromagnetic wave propagating in a plasma. The wave is
propagating along z-axis. In this moving frame the wave electric
field is given by
\begin{equation}
\mathbf{E}=\Omega A_{0}\left( \mathbf{e}_{x}\sin \Omega
t-\mathbf{e}_{y}\cos \Omega t\right).  \label{Etr}
\end{equation}
In the laboratory frame the wave dispersion equation is given by
\begin{equation}
\omega^{2}=k^{2}+\Omega^{2},  \label{DE}
\end{equation}
where $\Omega$ can be re-expressed in terms of the electron and
positron densities  $n = n_{0} /\gamma_{g}$ and of the particle
energy $(m^{2}+P^{2})^{1/2}$ in the laboratory frame  as $\Omega
=[(8\pi e^{2}n)/( m^{2}+P^{2})^{1/2}] ^{1/2}$. The  phase and
group velocities of a nonlinear electromagnetic wave in a plasma
depend  on the plasma parameters and on the wave amplitude. From
Eq.(\ref{DE}) we find that the phase velocity $v_{ph}=\omega /k$
and the group velocity $v_{g}=\partial \omega /\partial k$, are
related according to equation $v_{ph}v_{g}=1$. In the laboratory
frame the electric and the magnetic fields are given by
\begin{equation}
\mathbf{E}=-\partial _{t}\mathbf{A}=\omega
A_{0}[\mathbf{e}_{x}\sin (\omega t^\prime -kx)-\mathbf{e}_{y}\cos
(\omega t^\prime -kx)],
\end{equation}
\begin{equation}
\mathbf{B}=\nabla \times \mathbf{A}=kA_{0}[\mathbf{e}_{x}\cos
(\omega t^\prime -kx)-\mathbf{e}_{y}\sin (\omega t^\prime -kx)],
\end{equation}
respectively, with $t^\prime $  the time in the laboratory frame.

We see that in a plasma  the first invariant of the
electromagnetic field $\mathcal{F}= (E^2 - B^2)/2$ is not equal to zero and is
given by
\begin{equation}
\mathcal{F}= \frac{\Omega
^{2}}{2} A_{0}^{2}\equiv
\frac{1}{2}\left( \frac{\Omega }{\omega }\right) ^{2}E_{0}^{2}.
\end{equation}
It vanishes when the plasma density tends to zero, i. e., in
vacuum. In the following  we shall use the notation $E= \Omega
A_{0} \equiv (\Omega/\omega )E_{0}$.

\section{Kinetic description of the electron-positron plasma}\label{KD}

We consider the propagation of a circularly polarized
electromagnetic wave in an underdense collisionless plasma in the
reference frame moving with the wave group velocity $v_g$. The
relativistic kinetic equation
\begin{equation}
\frac{\partial f_{\alpha }}{\partial t}+e_{\alpha }\mathbf{E}\frac{\partial
f_{\alpha }}{\partial \mathbf{p}}=q_{\alpha }(E,p),  \label{kinetic}
\end{equation}
describes the dependence on time and momentum of the distribution
function $f_{\alpha }(\mathbf{p},t)$ in this moving frame where  a
spatially homogeneous electric field $\mathbf{E}$ is present. The
source term in Eq.(\ref{kinetic}) is proportional to the
quasiclassical probability
\begin{equation}
\exp \left[ -\frac{\pi (m^{2}+p_{\bot }^{2})}{|e\mathbf{E}(t)|}\right] .
\label{tun}
\end{equation} of tunneling through the gap between the lower and
the upper continuum of electron energy spectrum in the presence of
the electric field. We note that the form of Eq.(\ref{tun})
corresponds to the case of constant field \cite {el-pos}. However,
the characteristic time of pair production $c/l_{c}$, where
$l_{c}=\hbar /mc$ is electron Compton wavelength, is negligible
with respect to the wave period. This estimate gives a lower bound
on the pair production time while the estimate that follows from
the quasiclassical approximation gives for the time of sub-barrier
motion $t_{tun}=1/a\omega $ \cite{PopovMarinov,ssbul}, where
$a=eA/mc$ is the dimensionless amplitude of vector-potential.
However, for $a\gg 1$ (the case we are considering), even this
estimate  yields a pair production time much shorter than the wave
period. Thus, it is possible to use Eq.(\ref{tun}) for the
time-varying electric field with time playing the role of a
parameter. In addition, following the reasoning of Refs. \cite{Matsui, Matsui1,
Kluger}, we assume that the pairs are produced at
Rest, i. e. the momentum distribution of the source term is
taken to be proportional to the Dirac delta function
\begin{equation}
q_{\alpha }(E,p)=2 e^2\mathbf{E}^2(t)\exp \left[ -\frac{\pi
m^{2}}{|e\mathbf{E}(t)|}\right] \delta \left( \mathbf{p}\right) .
\label{source}
\end{equation}
This assumption is reinforced by the fact that the momentum
distribution in Eq.(\ref{tun}) has a width  $p_{\bot }\sim
(|e\mathbf{E}(t)|)^{1/2}= m(|\mathbf{e}(t)|)^{1/2}\ll m$  which
is negligible with respect to the momentum that electrons
(positrons) acquire in the electric field. Here
$\mathbf{e}=e\mathbf{E}/m^ 2=\mathbf{E}/E_{Sch}\ll 1$ is the
normalized electric field and $E_{sch}=m^2/e$ is the critical
Schwinger field. The source term has been normalized in such
way that $\int q_{\alpha }(E,p)d^{3}~p/(2\pi )^{3}$ gives the
number $[(eE)^{2}/4\pi ^{3}]\exp [-\pi m^{2}/eE]$ of positrons
(electrons) produced according to Schwinger's formula.

We solve Eq.(\ref{kinetic}) by integrating it  along the particle
characteristics. The equations for the characteristics for each
species $\alpha $ are
\begin{equation}
{p}_{\Vert} = {p}_{\Vert,0}, \qquad \frac{dp_{\bot }}{dt}=e_{\alpha
}E,\qquad \frac{df_{\alpha
}}{dt}=q_{\alpha },
\end{equation}
Introducing the function
$\mathbf{A}(t)=-\int\limits_{0}^{t}\mathbf{E}ds$, we obtain
\begin{equation}
\mathbf{p}_{\bot }=e_{\alpha
}\int\limits_{0}^{t}\mathbf{E}ds+\mathbf{p}_{\bot 0}\quad {\rm i. e.,}
\quad \mathbf{p}_{\bot }+e_{\alpha }\mathbf{A}(t)=
\mathbf{p}_{\bot 0}.
\end{equation}
As a result the distribution function is given by the  following expression
\begin{equation}
f_{\alpha }=f_{\alpha ,0}[p_\Vert, \mathbf{p}_{\bot }+e_{\alpha
}\mathbf{A}(t)]+\int\limits_{0}^{t}q_{\alpha
}\{\mathbf{p}_{\bot }+e_{\alpha }[\mathbf{A}(t)-\mathbf{A}(t^{\prime
})],t^{\prime }\}dt^{\prime
},  \label{f}
\end{equation}
where $f_{\alpha ,0}(p_|{\Vert}, \mathbf{p}_{\bot })$ is the
distribution function of the initial plasma positrons (electrons)
before the passage of the electromagnetic wave. Let us assume that
at the initial time  $t=0$ the plasma is cold so that
\begin{equation}\label{f0}
f_{\alpha ,0}=n_{0}{(2\pi )^{3}}\delta (p_{\bot })\delta
(p_{\Vert}-p_{\Vert ,0}),
\end{equation}
where we recall that $p_{\Vert }$ is the component of the particle
momentum  parallel to the electromagnetic wave propagation  and
$p_{\Vert ,0}$ is its initial value which arises from the Lorentz
transformation  from the laboratory to the moving frame.

The modification of the kinetic equation given by the source term
in Eq.(\ref {source}) must also be accompanied by a change of the
source term in Maxwell equations. The electron-positron pair
production by a spatially homogeneous time-varying electric field
leads to the appearance of a time-dependent electric dipole which
generates a polarization current. Thus  the current density in
Eq.(\ref{el-field}) acquires an additional term with respect to
the situation when no pair production is present.  Then
Eq.(\ref{JJ}) reads as \cite{Kluger}
\begin{equation}
\frac{d\mathbf{E}}{dt}=-4\pi \mathbf{j}_{tot}=-4\pi \left(
\mathbf{j}_{cond}+ \mathbf{j}_{pol}\right) ,  \label{max}
\end{equation}
where the conduction current is
\begin{equation}
\mathbf{j}_{cond}(t)=e\sum\limits_{\alpha =+,-}\int f_{\alpha
}(\mathbf{p},t)
\frac{\mathbf{p}}{(m^{2}+{p}^{2})^{1/2}}\frac{d^{3}p}{(2\pi)^3},
\end{equation}
and the polarization current is \cite{Matsui} (see the Appendix for details)
\begin{equation}
\mathbf{j}_{pol}(t)=\frac{\mathbf{E}(t)}{|\mathbf{E}(t)|^{2}}\sum\limits_{\alpha
=+,-}\int q_{\alpha
}(\mathbf{p},t)(m^{2}+{p}^{2})^{1/2}\frac{d^{3}p}{(2\pi)^3}.
\end{equation}
Using the distribution function  (\ref{f}), we obtain  the
following expressions for the current densities
\begin{equation}
\mathbf{j}_{cond}(t)
=-2e^{2}n_{0}\frac{\mathbf{A}_{ }(t)}{[m^{2}+\mathbf{p}_{\Vert
0}^{2}+e^{2}{A}_{}^{2}(t)]^{1/2}}-2e^{2}\int
\limits_{0}^{t}\frac{\mathbf{A}(t)-\mathbf{A}(t^{\prime
})}{[m^{2}+e^{2}|\mathbf{A}(t)-\mathbf{A}(t^{\prime
})|^{2}]^{1/2}}\frac{|e\mathbf{E} (t^{\prime })|^{2}}{8\pi
^{3}}\exp \left[ -\frac{\pi m^{2}}{|e\mathbf{E} (t^{\prime
})|}\right] dt^{\prime },
\end{equation}
\begin{equation}
\mathbf{j}_{pol}(t)=\frac{e^{2}m}{2\pi ^{2}}\mathbf{E}(t)\exp \left[ -\frac{\pi
m^{2}}{|e\mathbf{E}(t)|}\right],
\end{equation}
where ${A}_{ } =|\mathbf{A}(t)|$. In performing  the momentum
space integration  we have used the fact  that the pairs are
produced with zero momentum. Inserting these expressions for the
current densities into  the r.h.s. of Eq.(\ref{max}) and using
the dimensionless vector-potential $\mathbf{a}=e\mathbf{A}/m$ and
the normalized electric field $\mathbf{e}=e \mathbf{E}/m^2$,  we
obtain the equation for the electric field evolution in the
presence of pair production
\begin{equation}
\left\{
\begin{array}{l}
\displaystyle \frac{d\mathbf{a}(t)}{dt}=-m \mathbf{e}(t), \\
\displaystyle m \frac{d\mathbf{e}(t)}{dt}=\omega
_{p}^{2}\frac{\mathbf{a}(t)}{[{1+\tilde{p}_{\Vert
0}^{2}+{a}^{2}(t)}]^{1/2}}+\frac{\kappa}{m}
\int\limits_{0}^{t}\frac{\mathbf{a}(t)-\mathbf{a}(t^{\prime
})}{[1+|\mathbf{a}(t)-\mathbf{a}(t^{\prime
})|^{2}]^{1/2}}\frac{|\mathbf{e}(t^{\prime})|^{2}}{8\pi ^{3}}\exp
\left[ -\frac{\pi }{|\mathbf{e}(t^{\prime})|}\right] dt^{\prime }
\\ \displaystyle ~~~~~~~~~~~~-\frac{em^{2}}{2\pi
^{2}}\mathbf{e}(t)\exp \left[ -\frac{\pi }{|\mathbf{e}(t)|}\right]
,
\end{array} \right.   \label{maxwell}
\end{equation}
where $\omega _{p}=\left(8\pi e^{2}n_{0}/m\right)^{1/2}$ is the
non-relativistic Langmuir frequency, $\tilde{p}_{\Vert 0}\equiv
p_{\Vert 0}/m$ and $\kappa =8\pi e^{2} m^4$, where the factor
$m^4$ stands for the inverse  of the invariant Compton 4-volume
$m^4=c/l_c^4\approx 0.14\times 10^{53}$ cm$^{-3}$ s$^{-1}$.
Similar equations were obtained in Ref.\cite{Kluger}, where
however there was no initial distribution function and a spatially
homogeneous electric field in vacuum was used, which is not a
solution of Maxwell's equations.

\begin{figure}[ht]
\begin{tabular}{ccc}
\epsfxsize5.5cm\epsffile{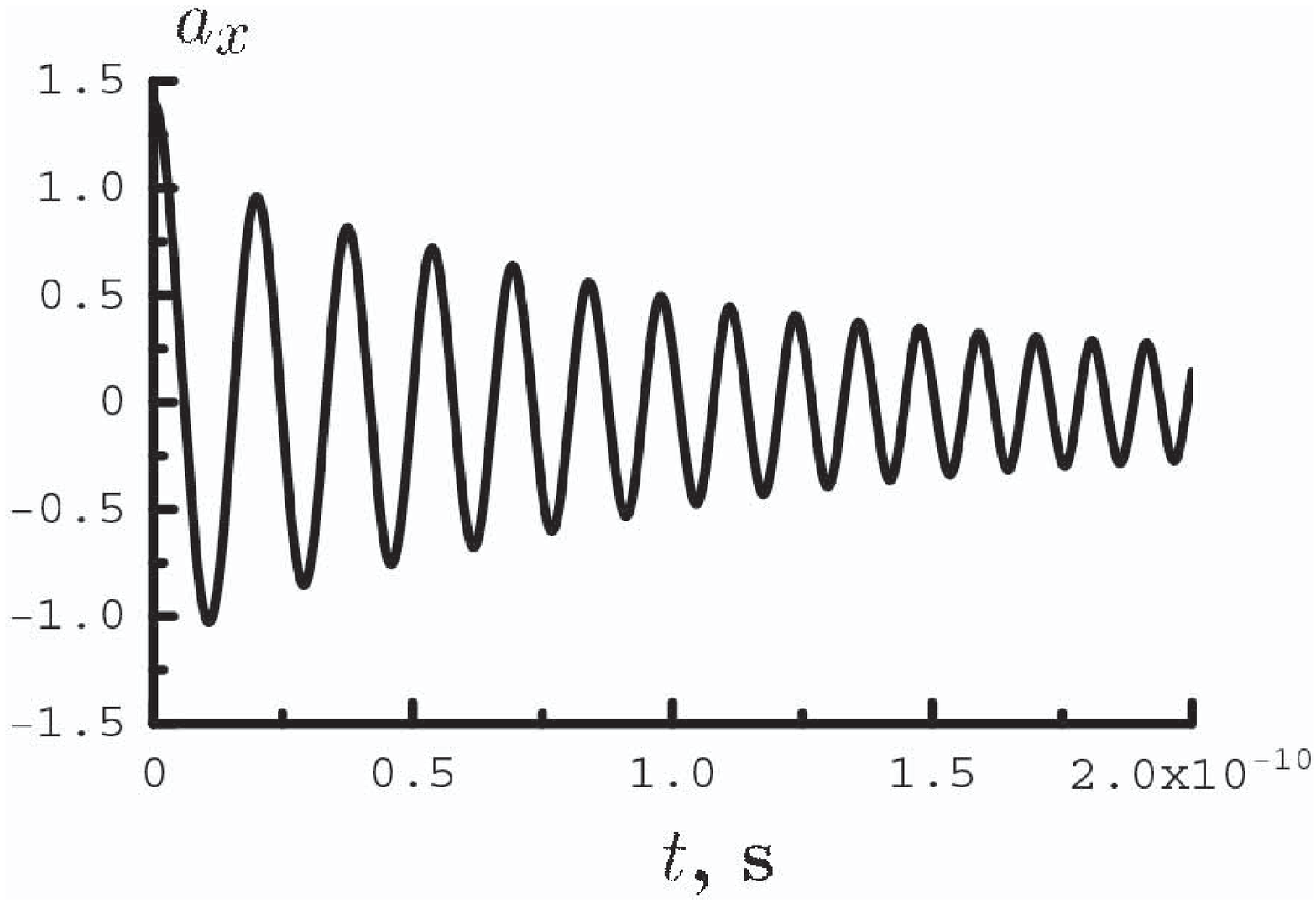} & \epsfxsize5.5cm\epsffile{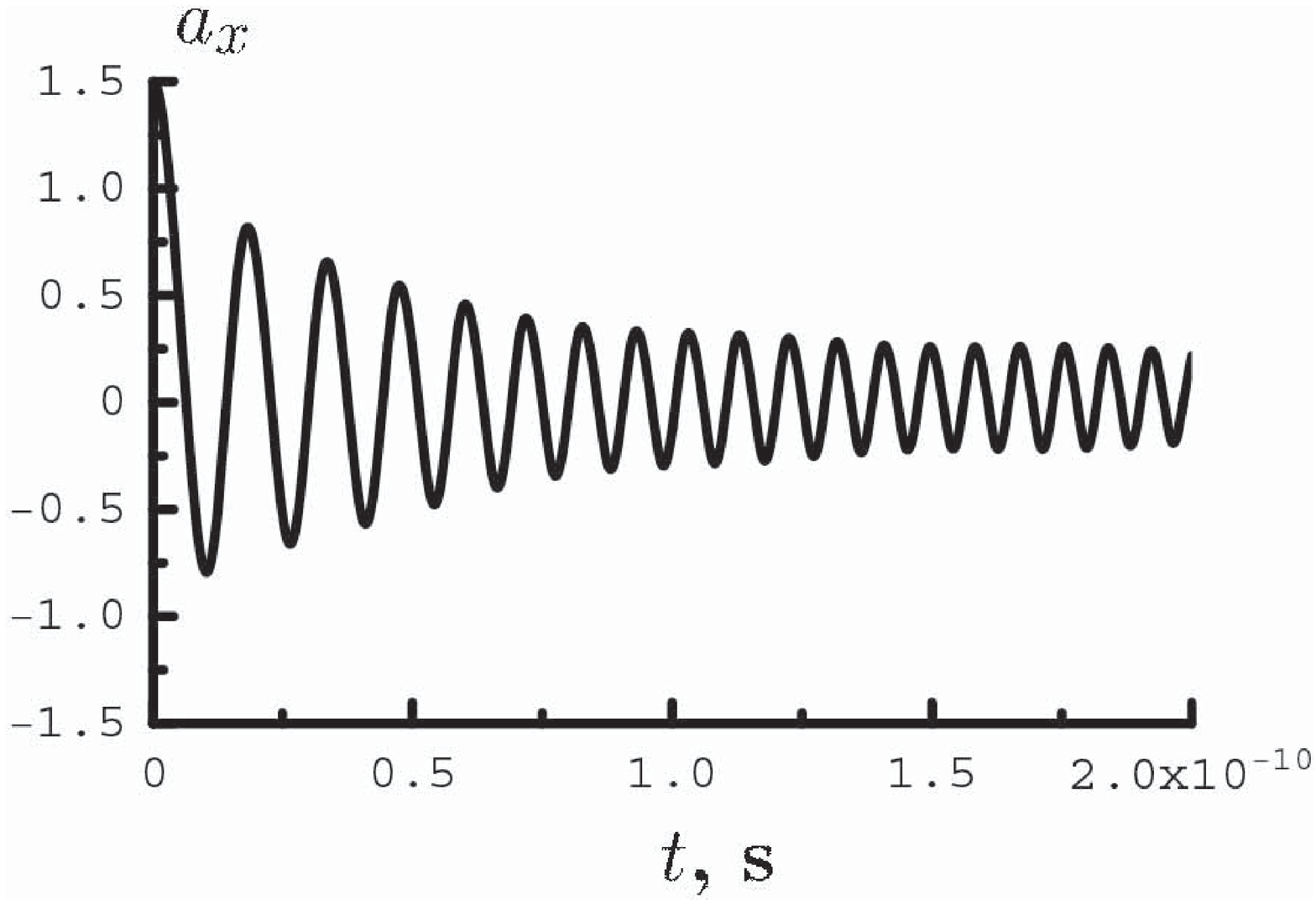} & %
\epsfxsize5.5cm\epsffile{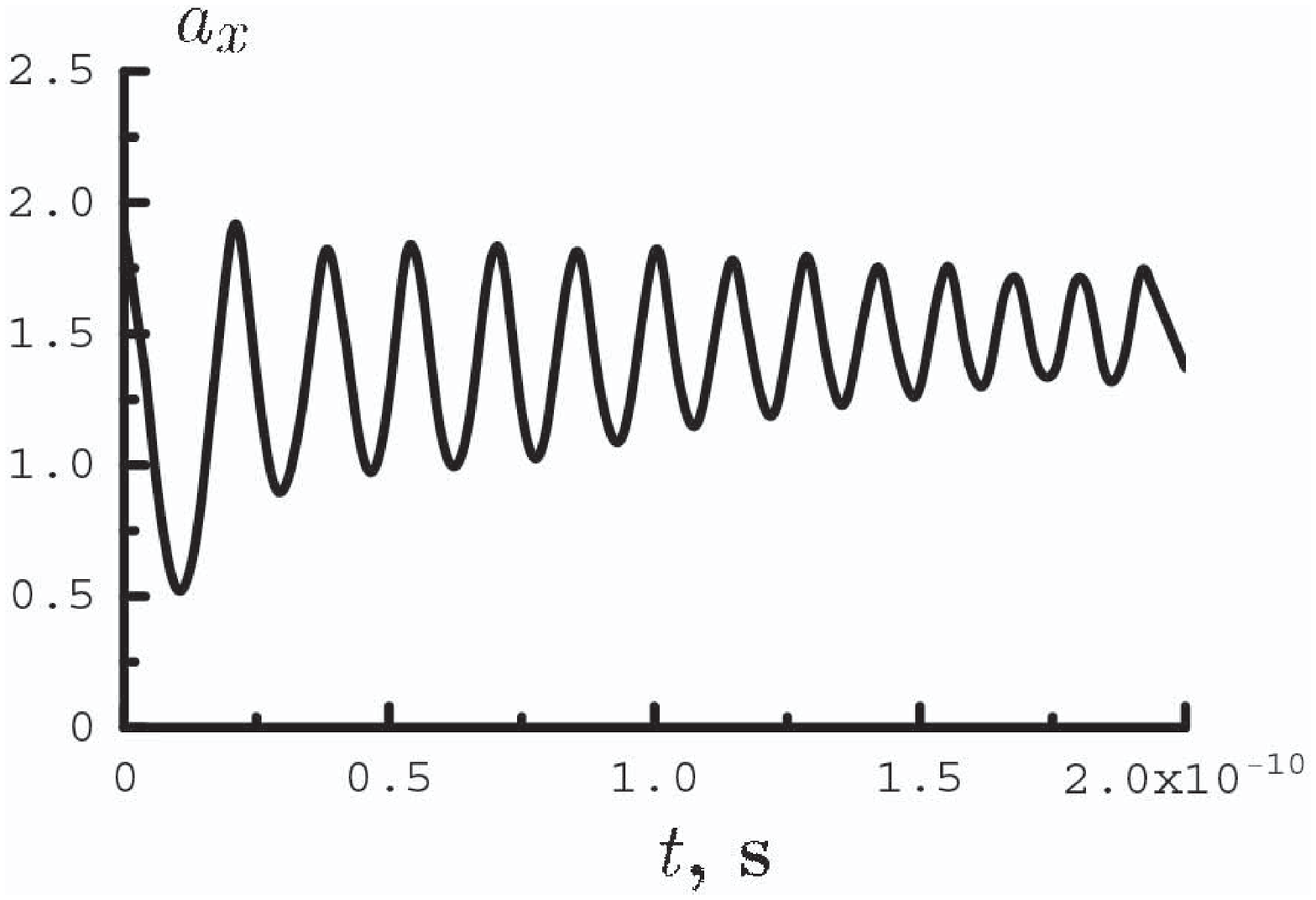} \\
\epsfxsize5.5cm\epsffile{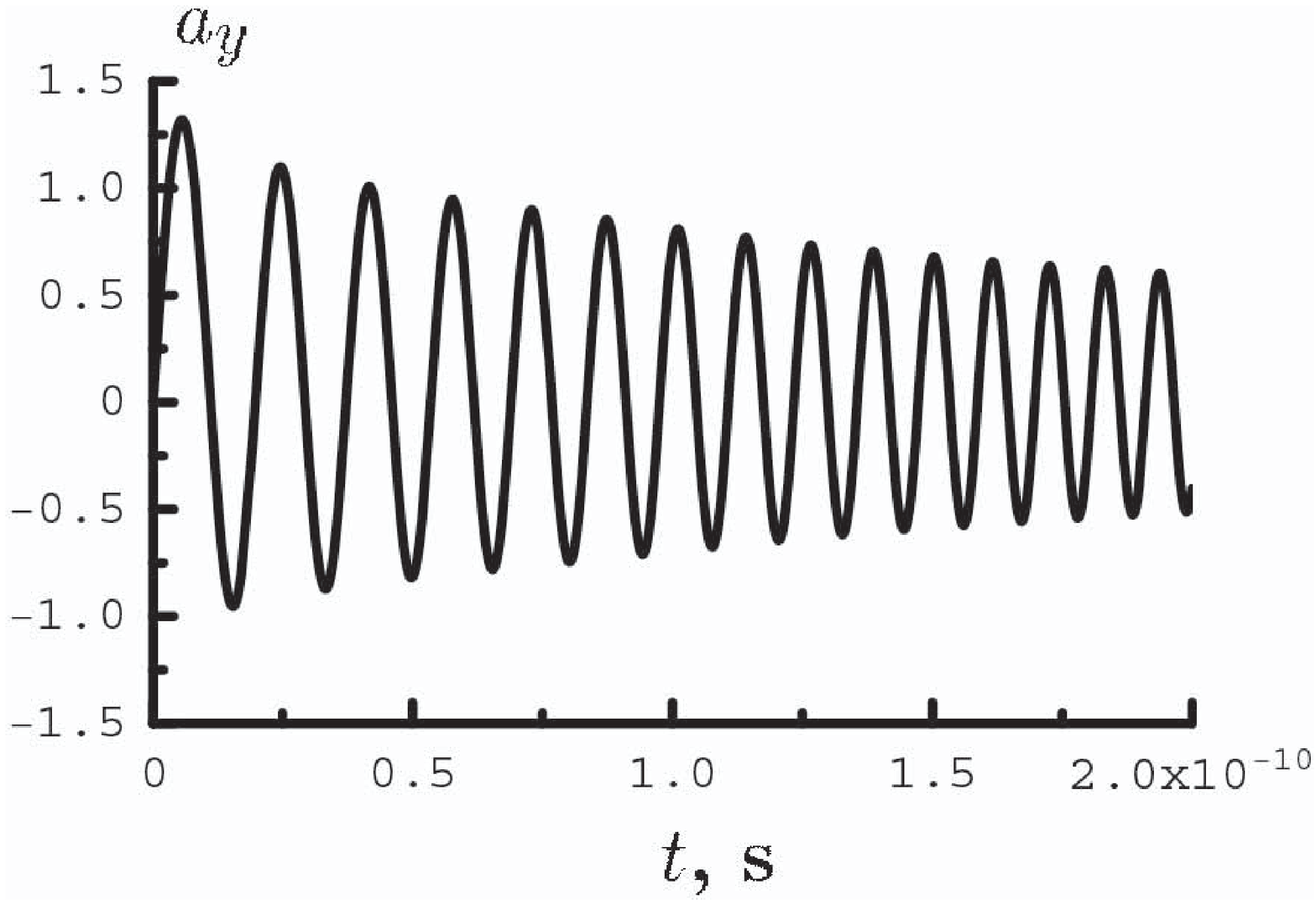} & \epsfxsize5.5cm\epsffile{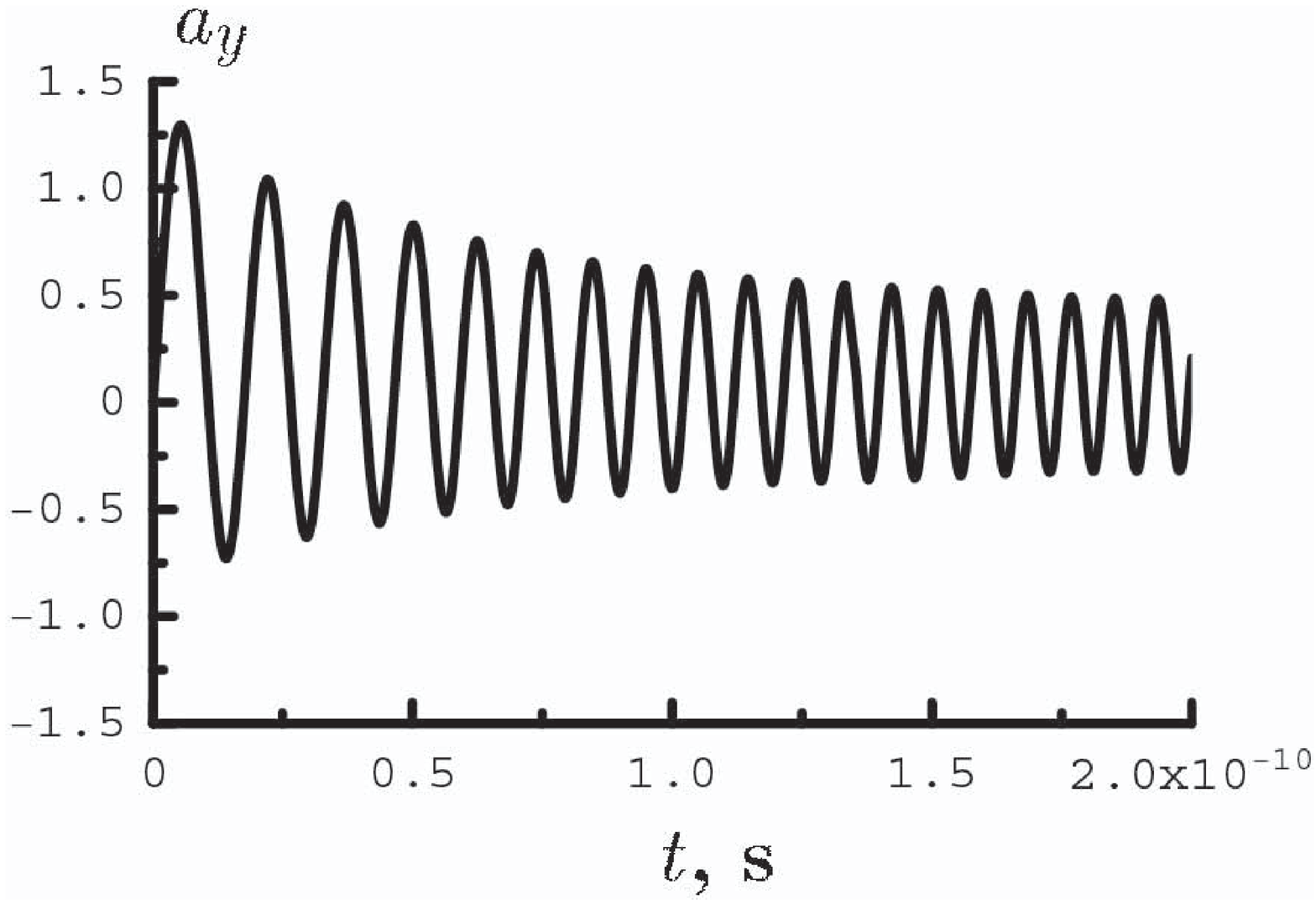} & %
\epsfxsize5.5cm\epsffile{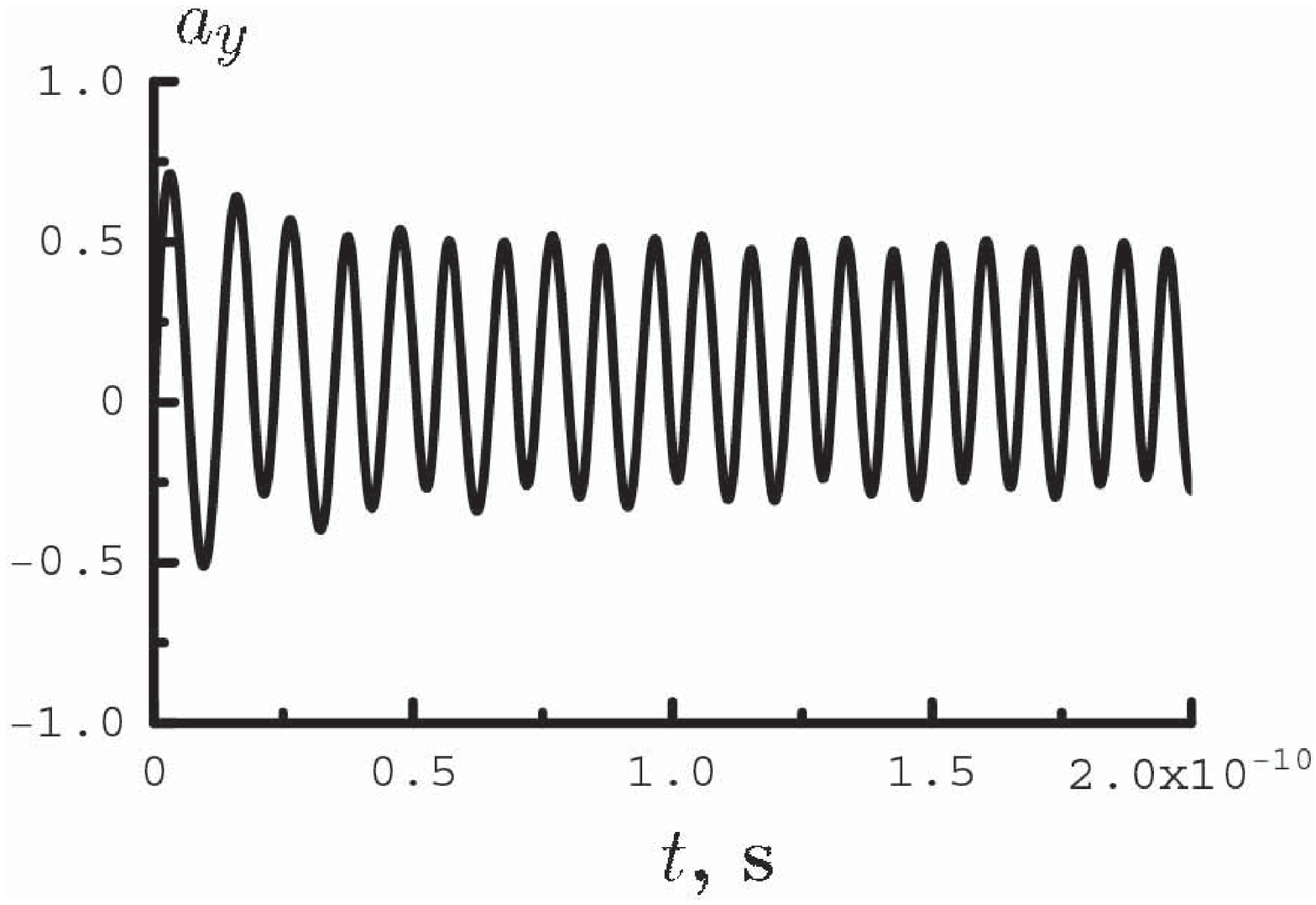} \\ a) & b) & c)
\end{tabular}
\caption{Time evolution in the moving frame of the $x$ and the
$y$-components of the dimensionless vector potential for different
initial amplitudes: $a=1.4\times 10^5$ (a), $a=1.5\times 10^5$
(b), $a=1.9\times 10^5$ (c) with initial plasma density
$n_0=10^{19}$ $cm^{-3}$ in the moving frame; $v_g\approx 1$,
$\gamma_g=10$. The upper row shows the $x$-component of the
vector-potential, the lower the $y$-component. On the $x$-axis
time is measured in seconds; $a=1$ corresponds, for a 1 $\mu$m
wavelength pulse,  to an intensity of $10^{18}$ W/cm$^2$ and
$a=4.6 \times 10^{5}$ to the Schwinger intensity.}
\end{figure}

The nonlinear integro-differential equation (\ref{maxwell}) cannot
be solved analytically. Numerical solutions of this equation are
presented in Fig.1 for different initial amplitudes. We can see
that the process  of electron-positron pair production  leads to
the damping of the wave  in the plasma and to the nonlinear
up-shift of its frequency. The damping is due to the fact that
each pair creation  takes a portion of the field energy equal to
$2mc^{2}$ as well as the amount needed for the particle
acceleration. The up-shift of the field frequency is due to the
increase of the plasma density, and thus of the Langmuir
frequency, as new pairs are created. This frequency up-shift is
seen in Fig.1, and bears a strong resemblance to the blue-shift of
an electromagnetic wave that propagates in a medium that becomes
ionized, as investigated theoretically in Ref.\cite{ionizationth}
and experimentally  in Ref.\cite{ionizationexp}.

Since the pair production rate depends on the field amplitude
exponentially, an unbalanced damping of the field components can
occur and lead to a change of the field polarization. We should
also note that the decrease of the amplitude of the
vector-potential is accompanied by a frequency up-shift, so that
the decrease of the amplitude of the  electric field  is not as
fast as that of the vector-potential. These properties of the
electric field are shown in Fig.2, where the projections of the
polarization vector  are presented for the same set of initial
parameters as in Fig.1. In Fig.2a we see the damping of the
$x$-component of the electric field and the transition from
circular to  elliptic polarization  with the major  axis of the
ellipse directed along the $y$-axis. In addition, in Fig.2b we see
a rotation of the principal axes of the ellipse.  The rotation of
the principal axes of the polarization ellipse was discussed in
Ref.\cite{far} in the case of the free propagation (without source
terms) of an elliptically polarized non liner pulse in a plasma.
The situation shown in Fig.2c is different from the two previous
ones. In this latter case the pair production rate at the
beginning of the field evolution is so large (see Fig. 3c) that
the first wave oscillation cycle cannot be completed, leading to
oscillations of the $x$-component of the wave vector potential
around a non-zero mean value determined by the balance between the
time averaged parts of the first two terms on the r.h.s. of the
second of Eqs. (\ref{maxwell}). This shift of the center of the
oscillations of the $x$-component of the vector potential leads to
a reduction of the  oscillation frequency of this wave component
so that, in this case, the $x$ and the $y$-components of the wave
oscillate at different frequencies.

\begin{figure}[ht]
\begin{tabular}{ccc}
\epsfxsize5.5cm\epsffile{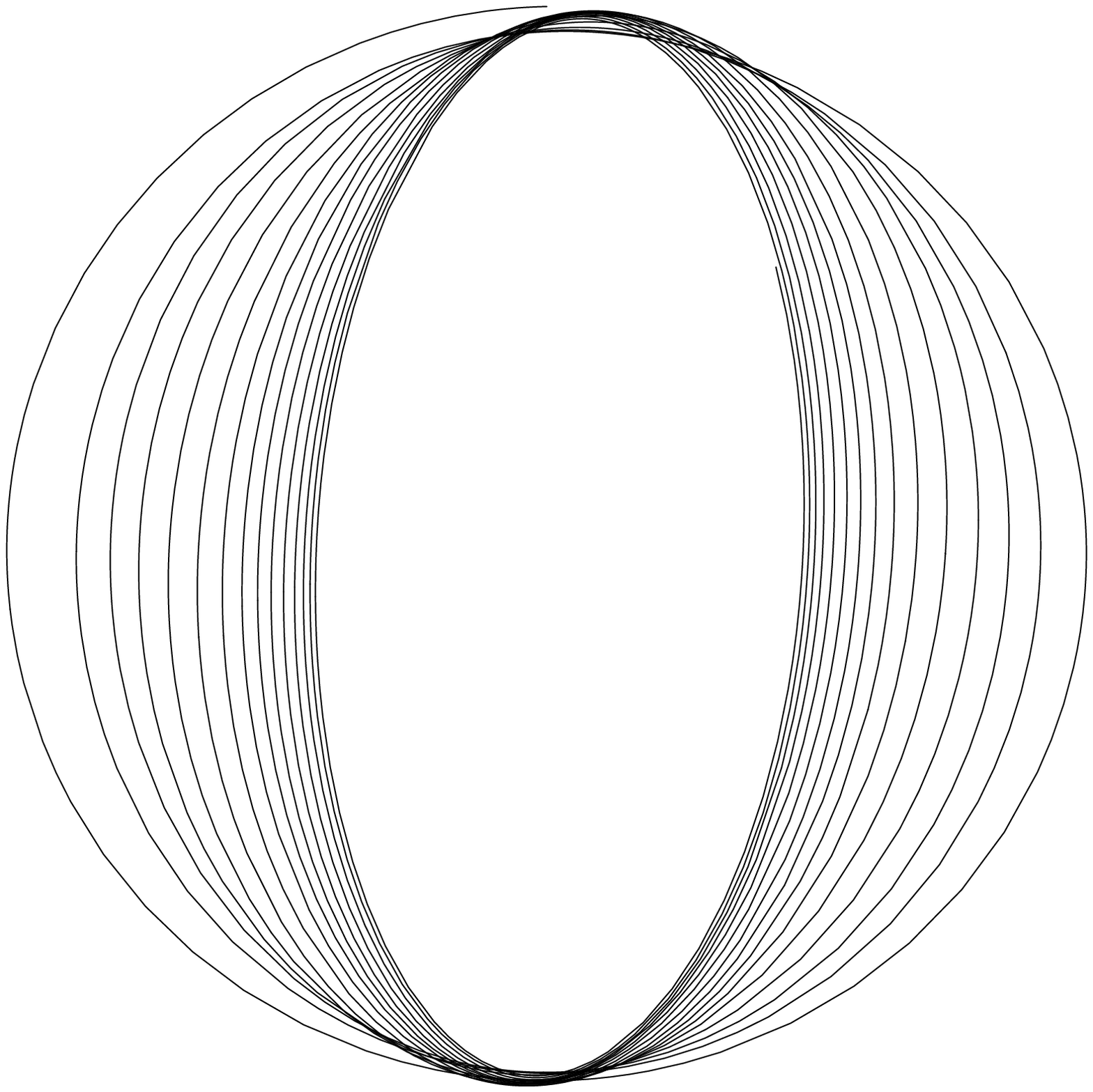} & \epsfxsize5.5cm\epsffile{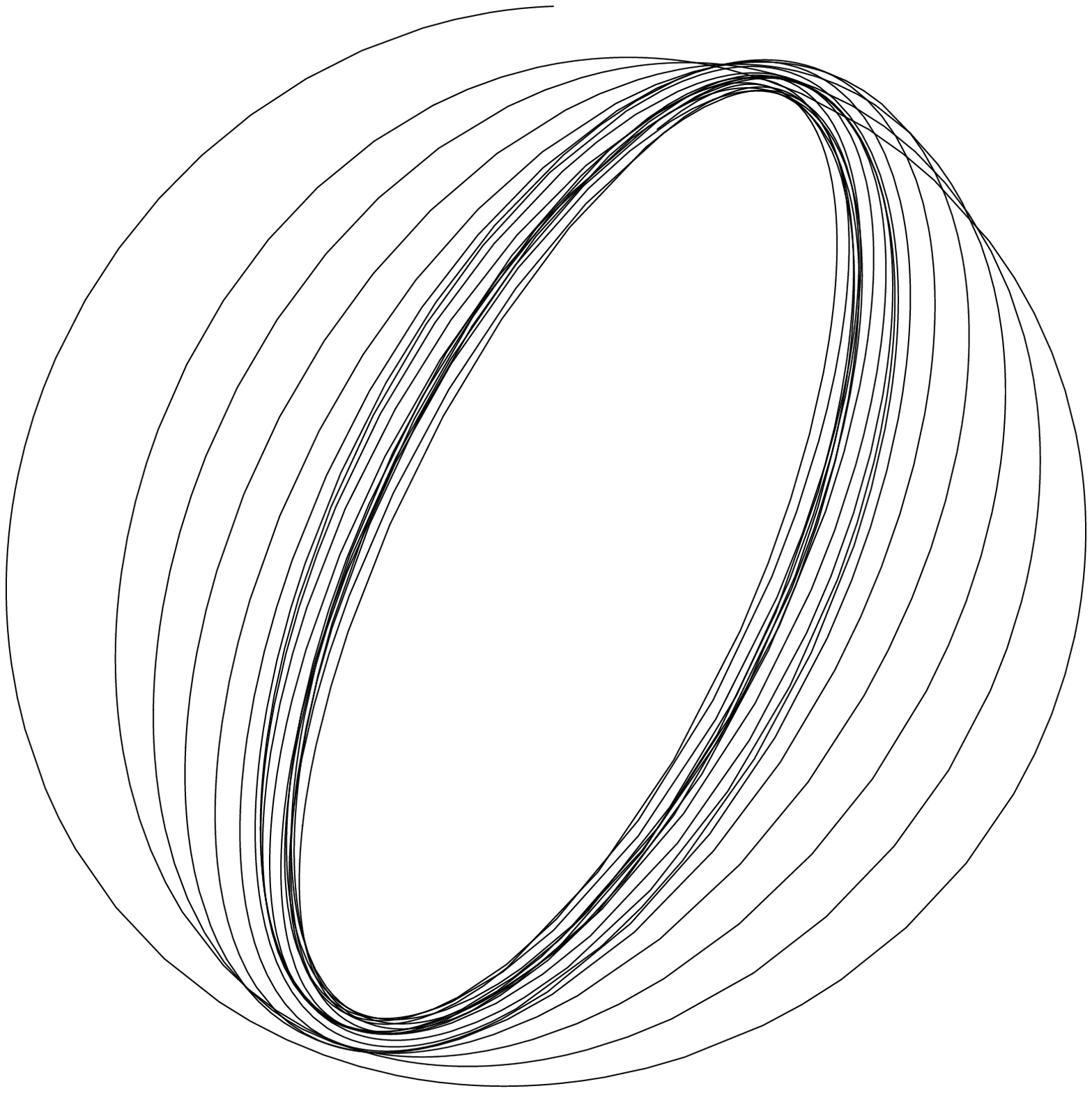} & %
\epsfxsize5.5cm\epsffile{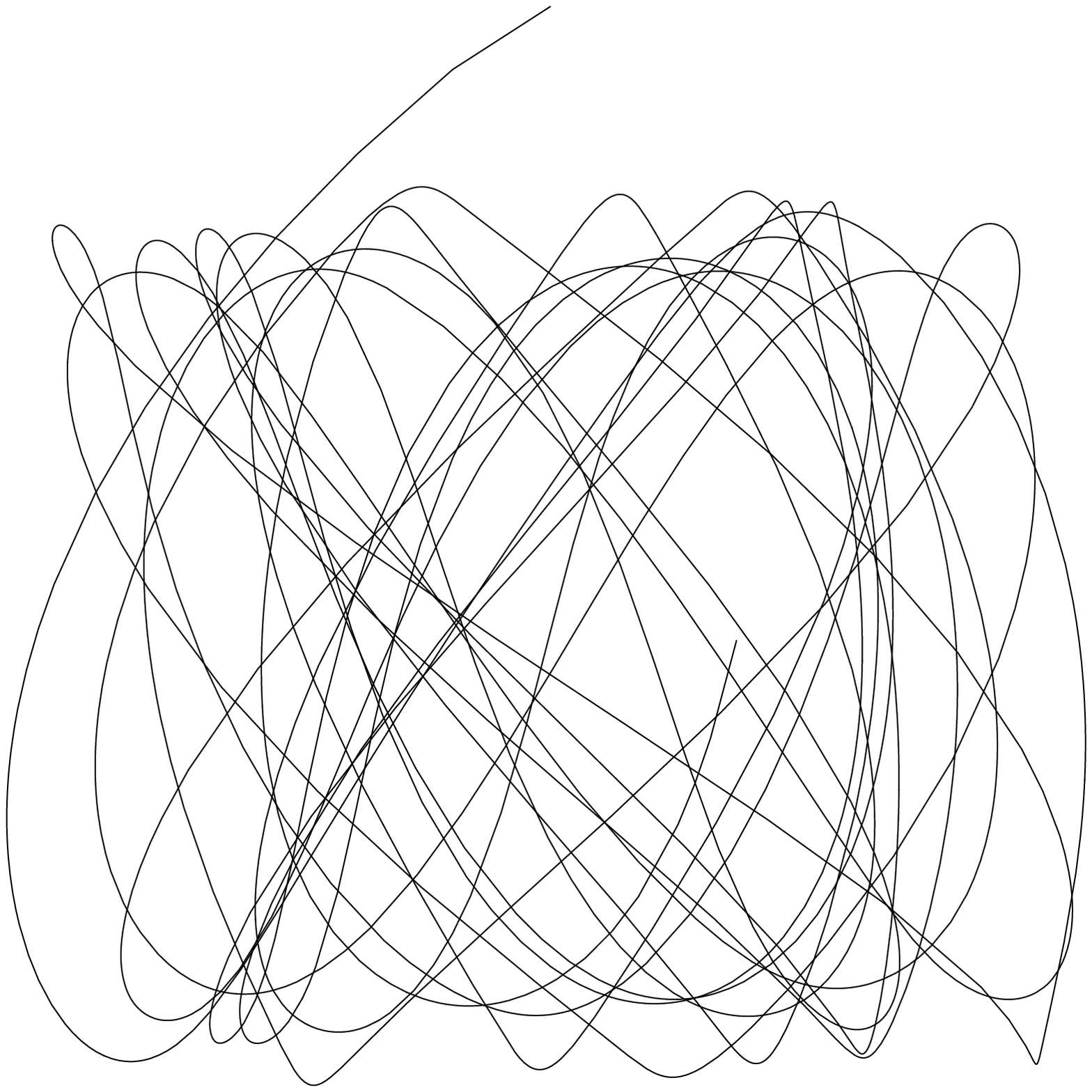} \\
a) & b) & c)
\end{tabular}
\caption{Trajectories of the projections of the electric field
polarization vector for the same set of initial conditions as in
Fig.1. }
\end{figure}

The increase with time of the pair plasma density is shown in
Fig.3 starting from an initial density before  the start of the
pulse propagation.  We  see that the particle density increases in
steps as a function of time. This time dependence is due to the
oscillations of the amplitude of the electric field  that  occurs
in an elliptically polarized wave (see Fig.2). The
electron-positron pairs are mainly produced  near the maxima of
the electric field amplitude, while there is practically no pair
production in between.   From these plots we can also see  that
the frequency is up-shifted since the width of the steps decreases
with time. For the pulse amplitude of frame c, the  pair production
occurs almost instantaneously at the beginning of the pulse
evolution.

\begin{figure}[ht]
\begin{tabular}{ccc}
\epsfxsize5.5cm\epsffile{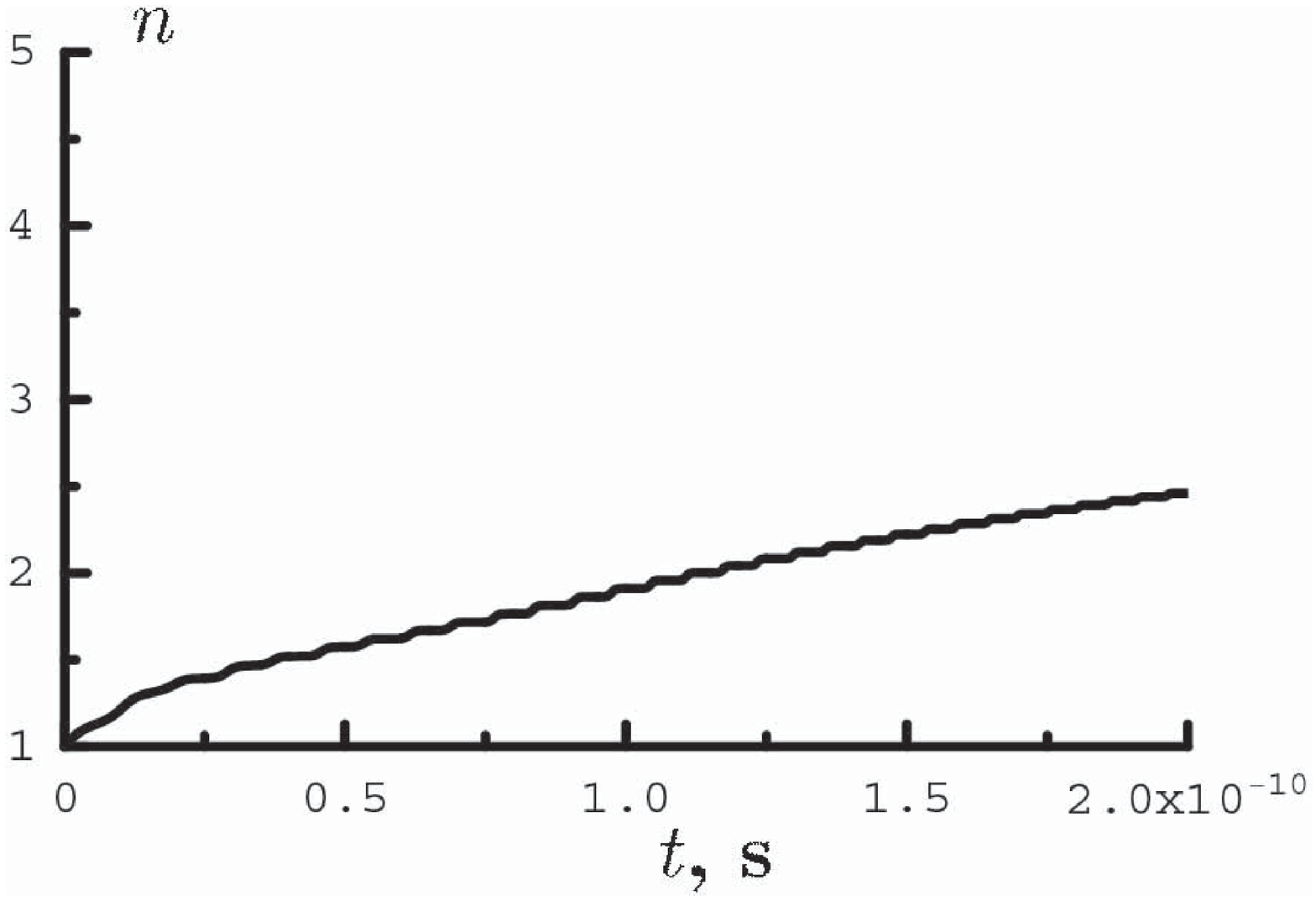} & \epsfxsize5.5cm\epsffile{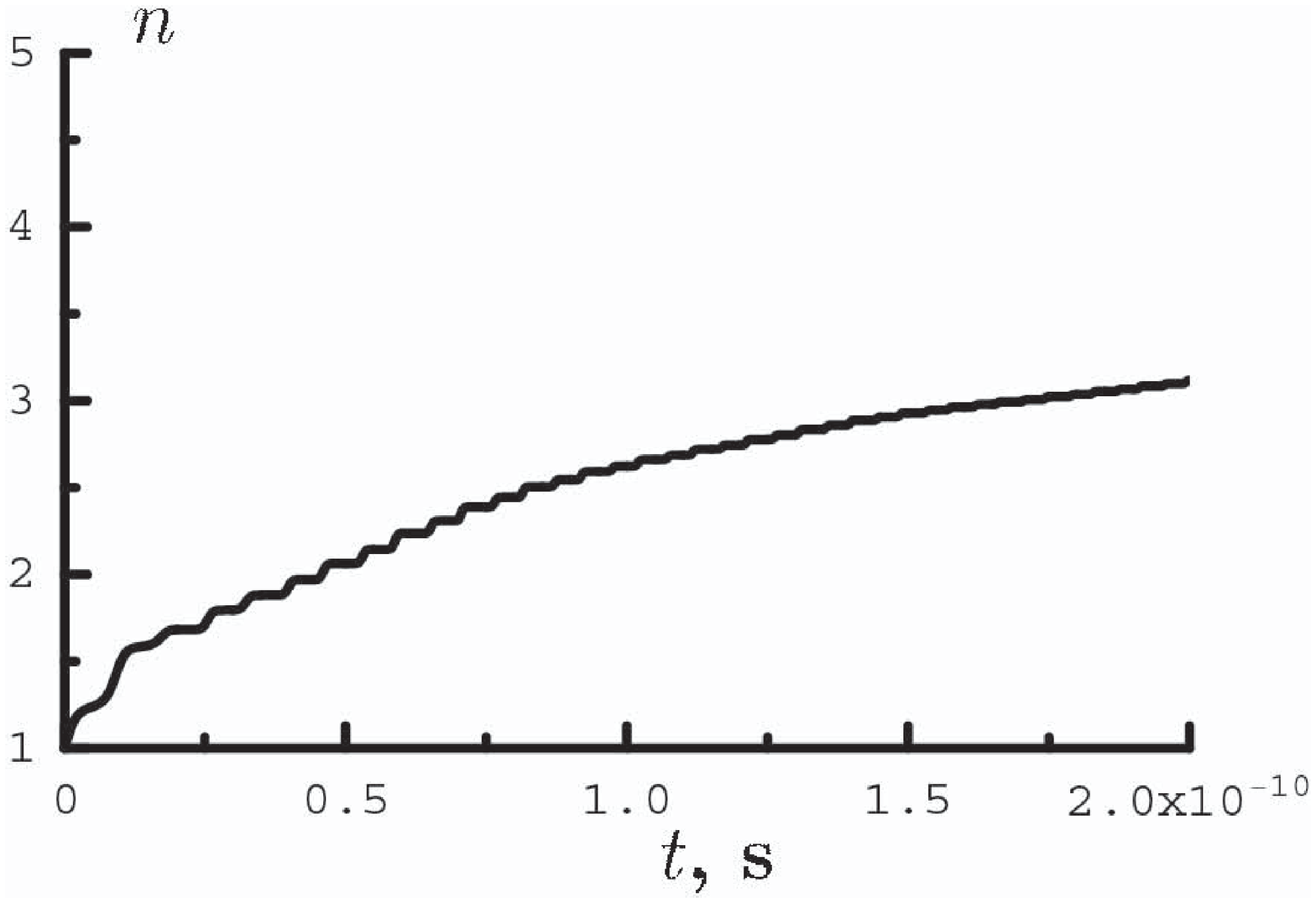} & %
\epsfxsize5.5cm\epsffile{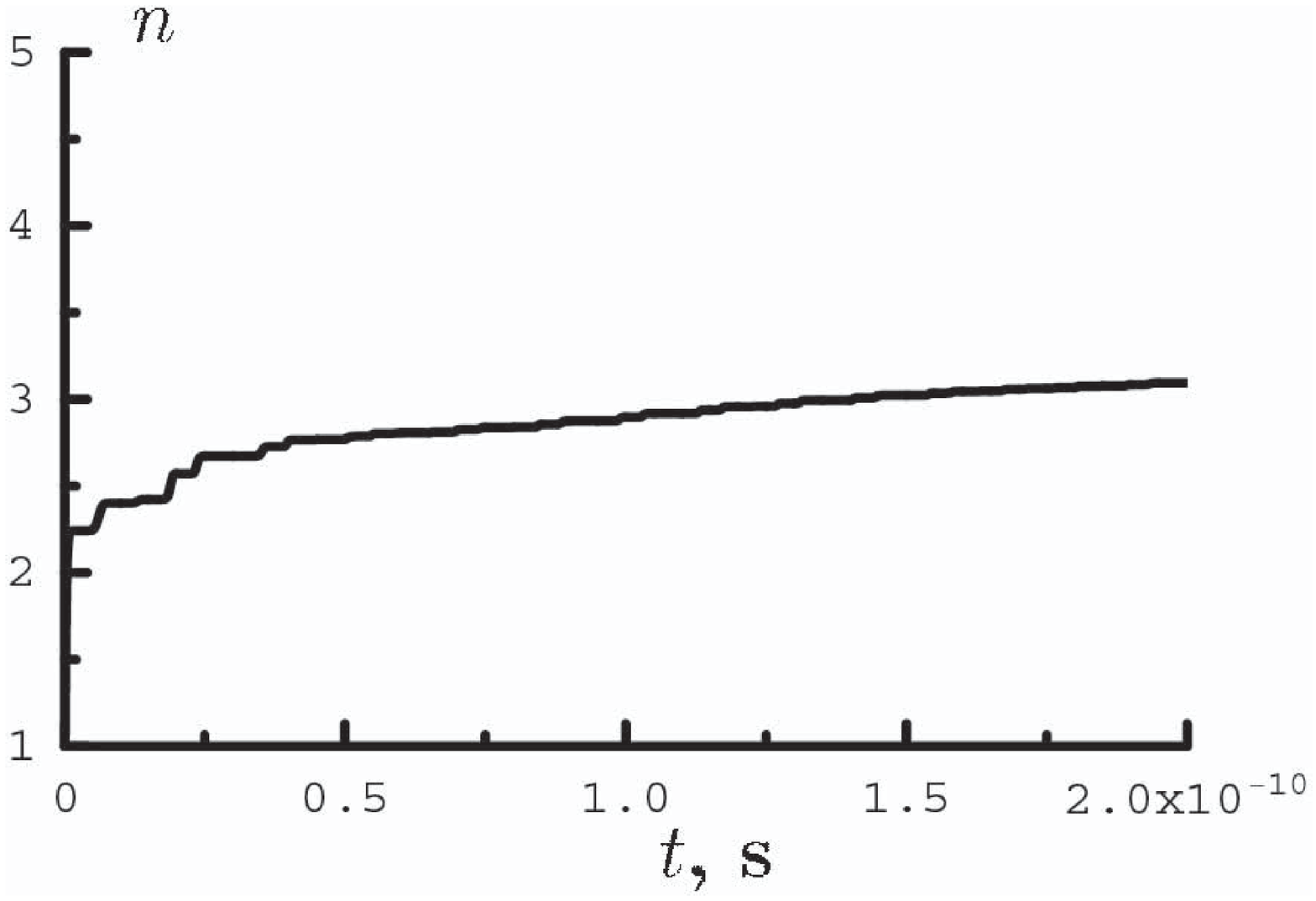} \\
a) & b) & c)
\end{tabular}
\caption{Dependence of the  plasma electron (positron) density on
time for the same set of initial conditions as in Fig.1 in the
moving frame. The density is measured in units of $10^{19}
cm^{-3}$ and time in seconds.}
\end{figure}

The difference between the above three cases is clearly
illustrated by the different shapes of the  particle distribution
functions in the $p_x$-$p_y$ plane (Note that the electron and the
positron distributions  are one the mirror image of the other). In
cases a) and b)  electrons and positrons are mostly created at the
maxima of the electric field $|{\mathbf E} |$ (and thus of the
vector potential $|{\mathbf A}|$). Since  at birth ${\mathbf
p}_\perp =0$, in the case of a circularly polarized electric field
this should lead to a ring type distribution. However, since the
wave polarization becomes   elliptical because of the backreaction
due to the pair creation, the distribution function of each
population consists, in the canonical momentum ${\mathbf p}_\perp
+ e_a {\mathbf A}$ plane, mainly  of two blobs at $\pm  e_a
{\mathbf A}_{\max}$. In the $p_x$-$p_y$  plane, these blobs  move
according to the time evolution of the vector potential ${\mathbf
A}_{}$. On the contrary the position of the initial distribution
function (denoted by a dark dot in the figure) corresponds to
${\mathbf p}_\perp + e_a {\mathbf A} = 0$. In case c) the pairs
are created mostly at the start at ${\mathbf p}_\perp + e_a
{\mathbf A} = e_a {\mathbf A}(t=0)$.  Since the time evolution of
${\mathbf A}(t)$  is ergodic, as shown in Fig.2,  their
distribution tends to be randomized in the $p_x$-$p_y$  plane.

\begin{figure}[ht]
\begin{tabular}{ccc}
\epsfxsize5.5cm\epsffile{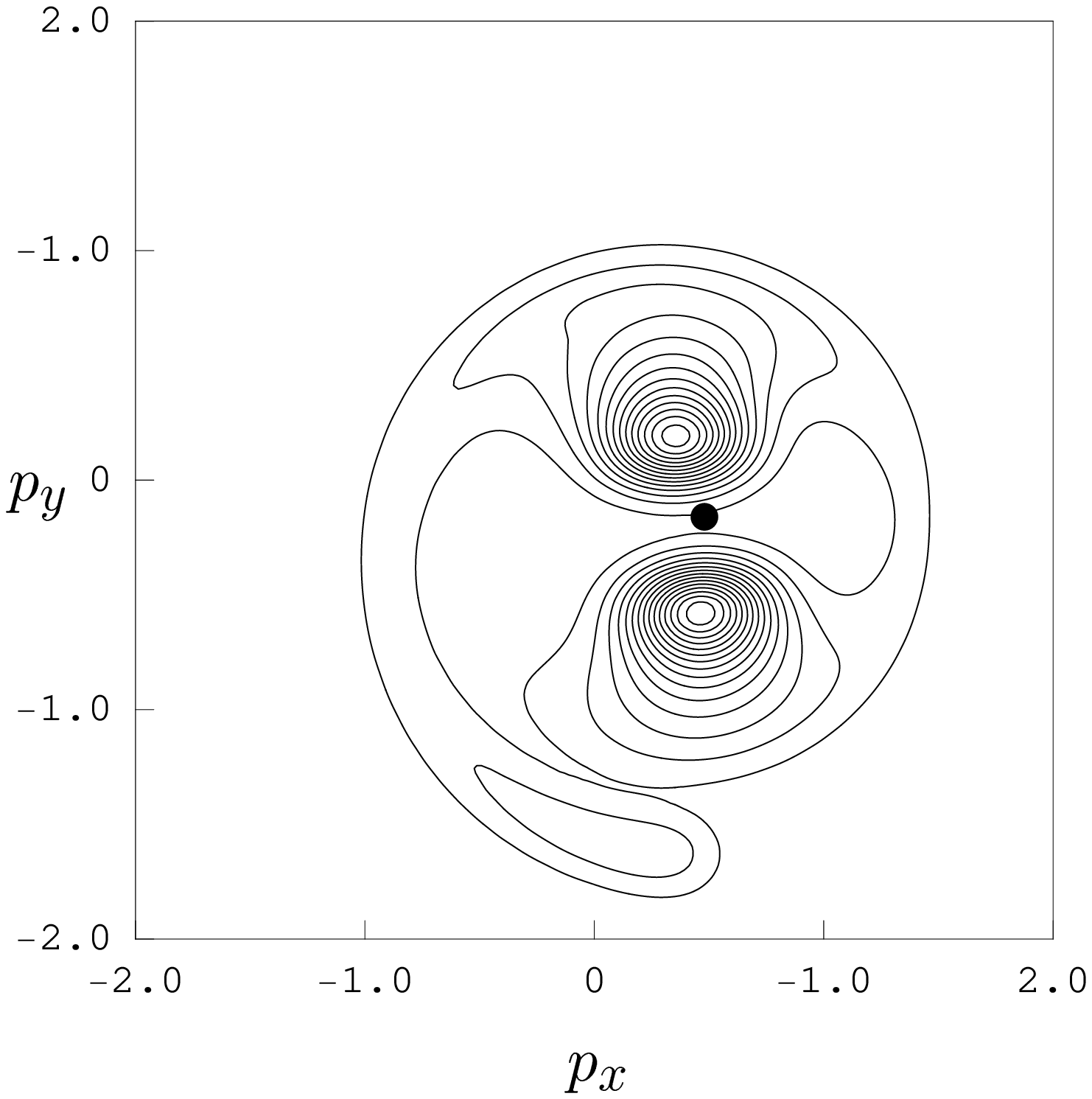} & \epsfxsize5.5cm\epsffile{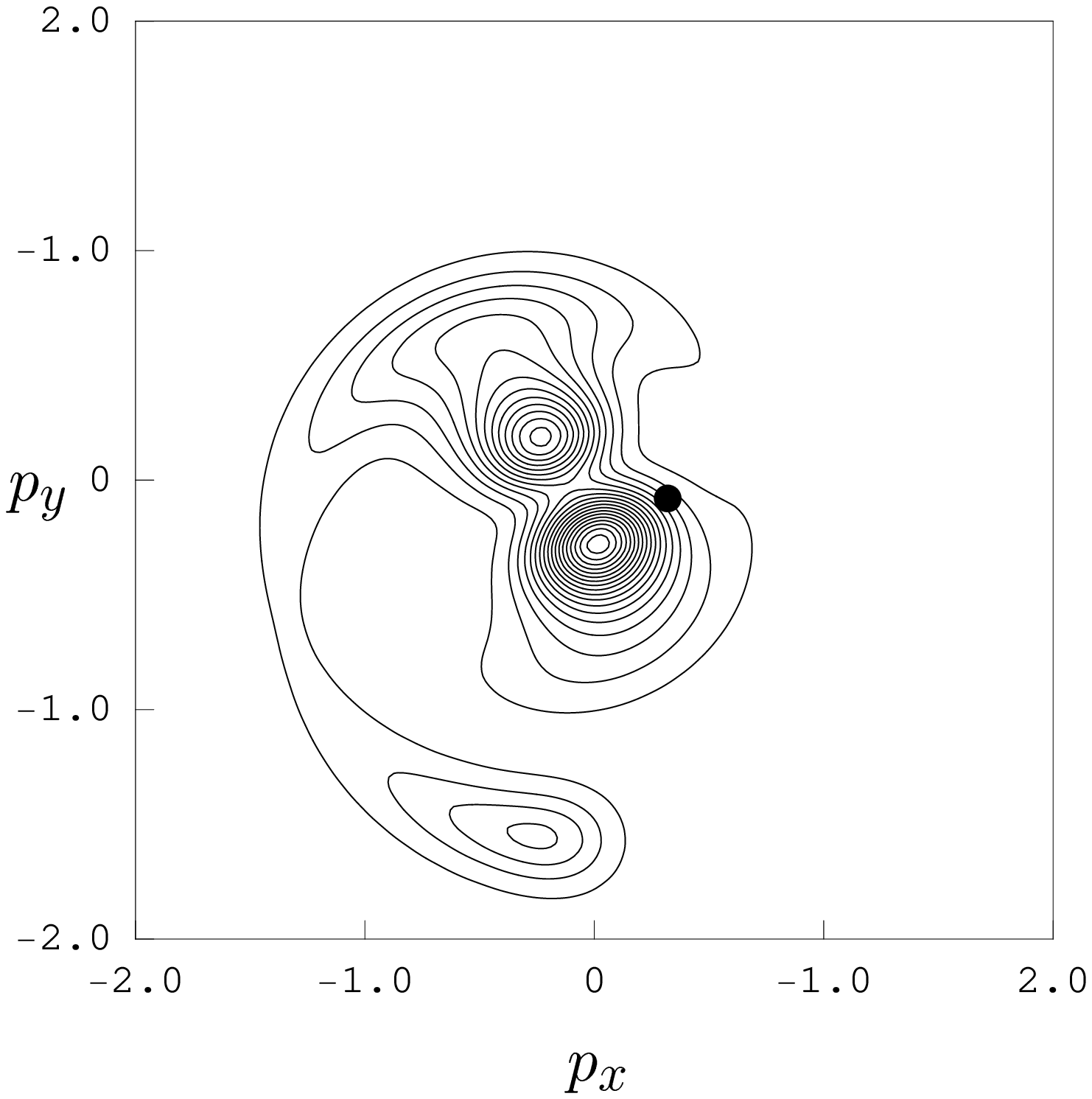} & %
\epsfxsize5.5cm\epsffile{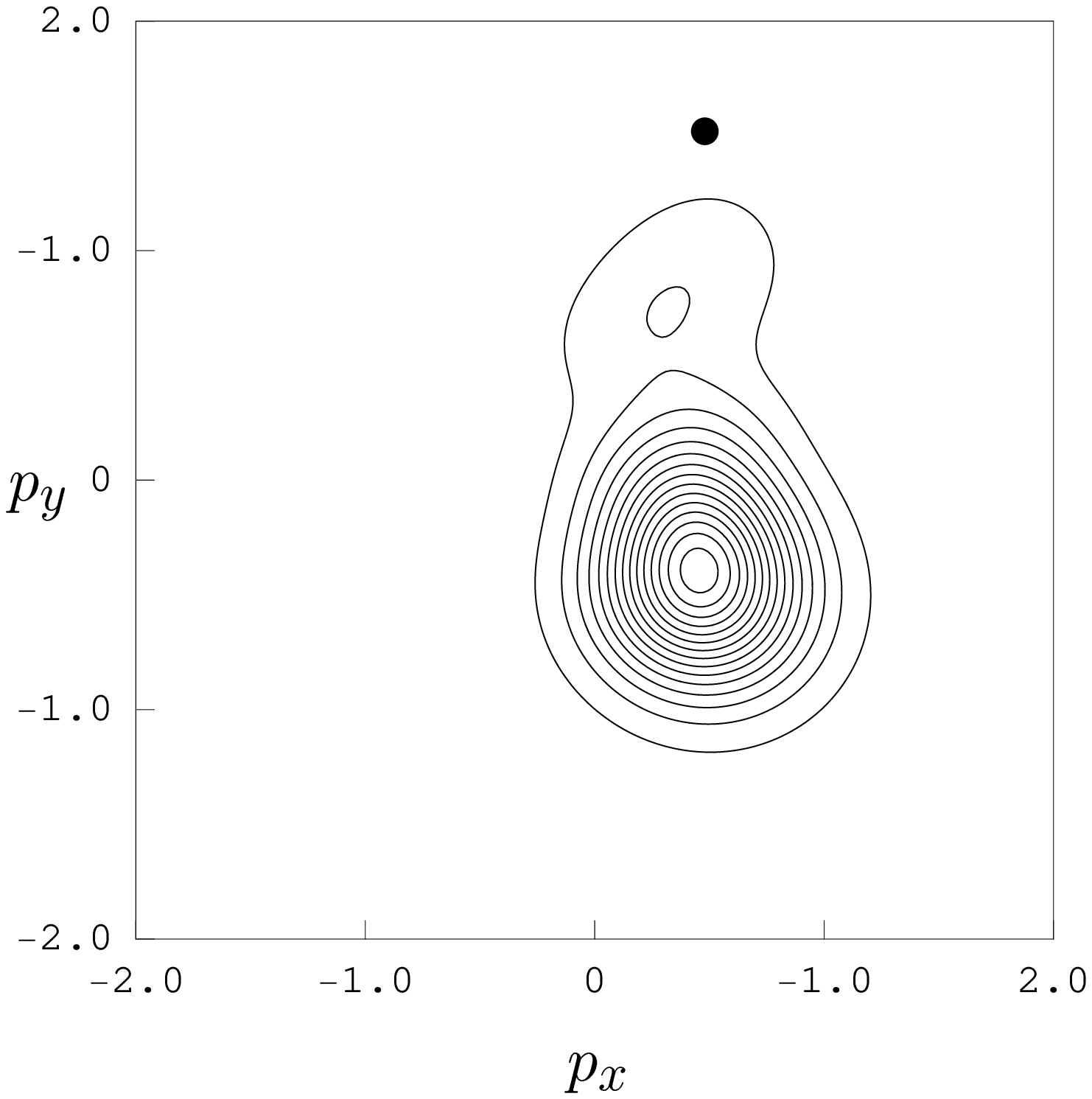} \\ a) & b) & c)
\end{tabular}
\caption{The electron distribution functions versus $p_x$ and
$p_y$  in the moving reference frame for the same set of initial
conditions as in Fig. 1 at time $2\times 10^{-10}$ s.  Particle
momenta are normalized on the dimensionless vector-potential
amplitude $a$ multiplied times $10^5$. Black circles correspond to
the initial plasma particle distribution at time $2\times
10^{-10}$ s.}
\end{figure}

\begin{figure}[ht]
\begin{tabular}{ccc}
\epsfxsize5.5cm\epsffile{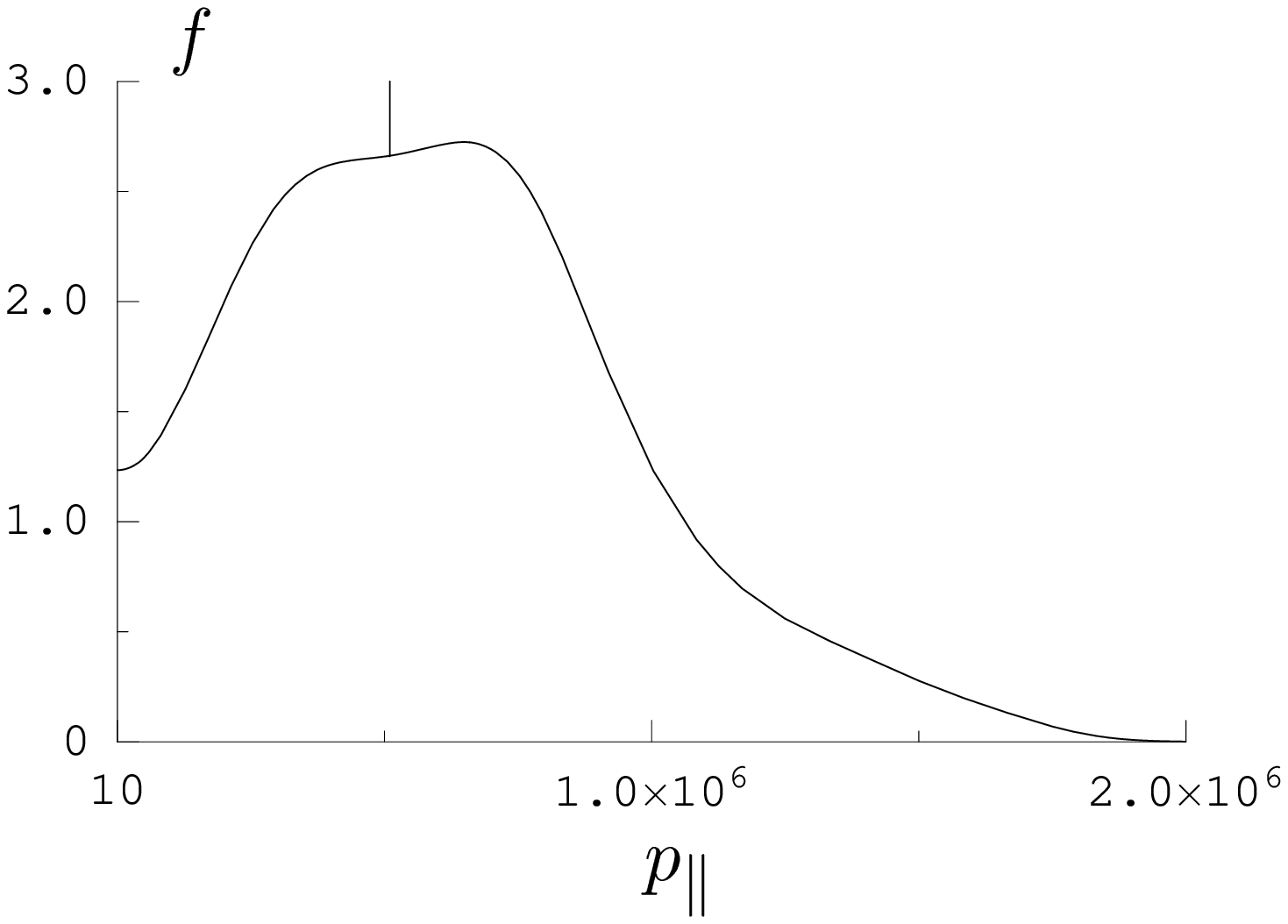} & \epsfxsize5.5cm\epsffile{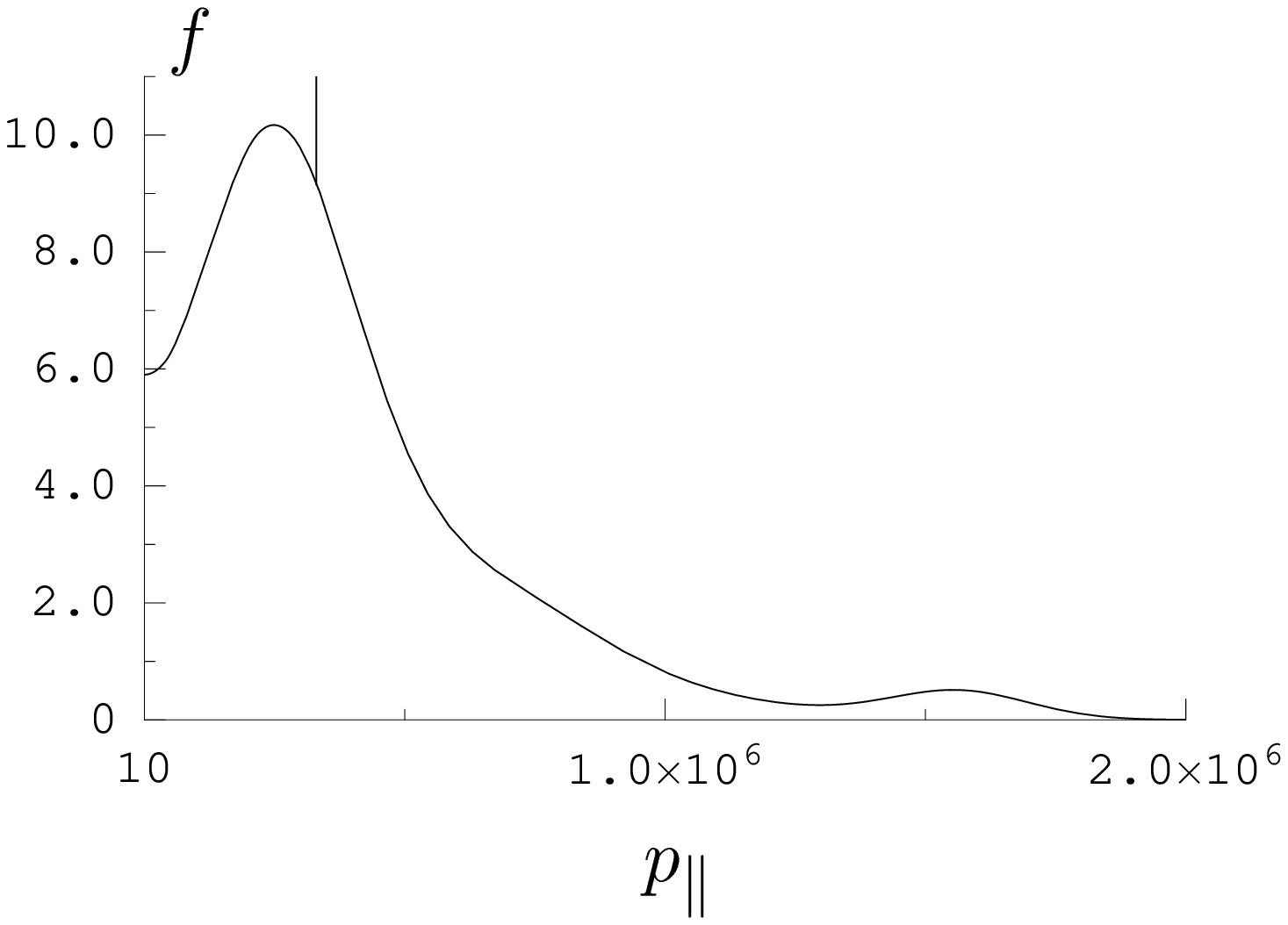} & %
\epsfxsize5.5cm\epsffile{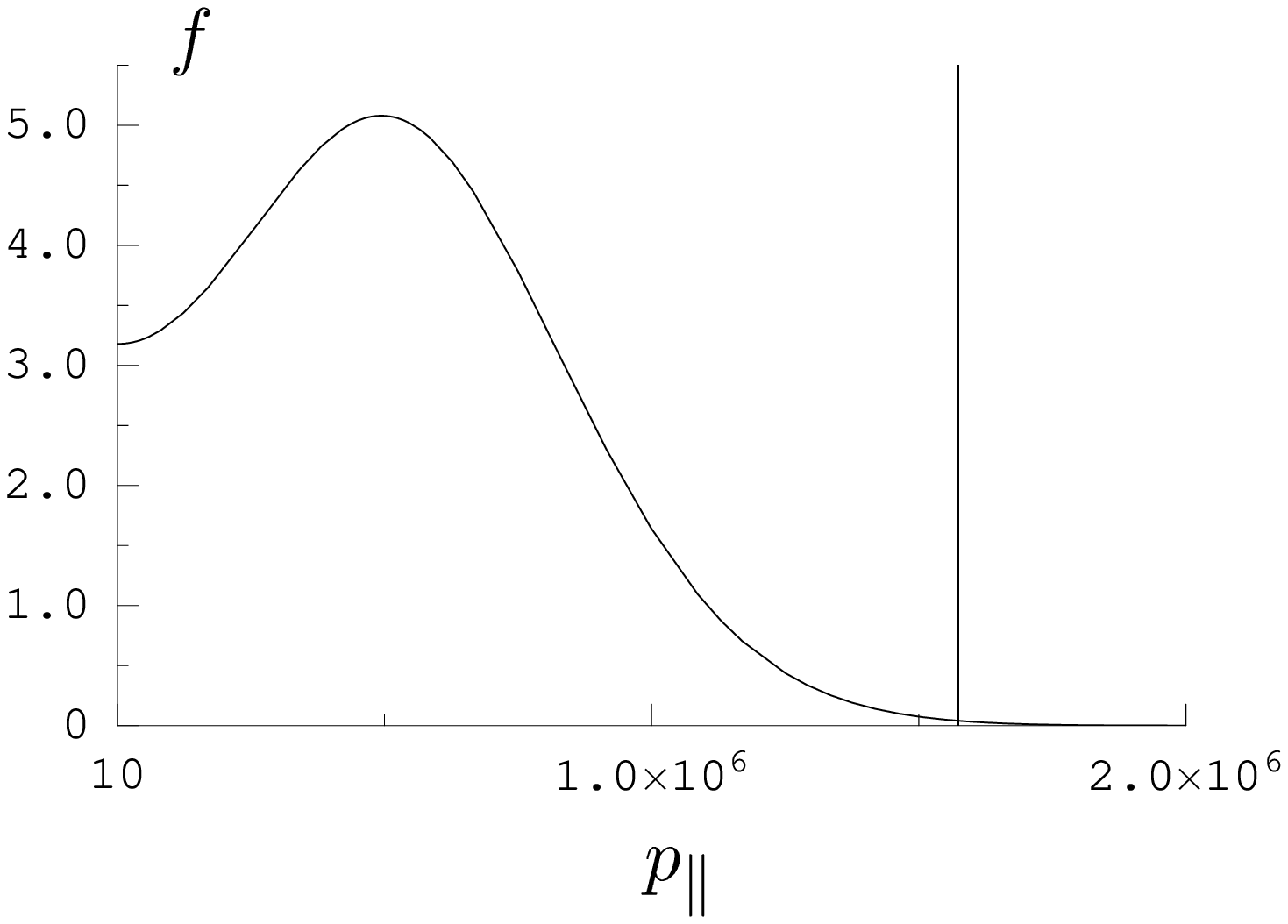} \\ a) & b) & c)
\end{tabular}
\caption{The electron distribution functions versus $p_\|$ in the
laboratory reference frame for the same set of initial conditions
as in Fig. 1 at time $2\times 10^{-10}$ s. Momentum is measured in
the same units as in Fig. 4. The vertical lines correspond to the
distribution of the initial plasma particles.}
\end{figure}

The  particle distribution function is shown in Fig. 5 versus the
parallel momentum $p_\|$ in the laboratory frame  at time
$t=2\times 10^{-10}$ s. Note that in case c) the strong damping of
the wave due to the pair creation  and the resulting non adiabatic
interaction has lead  to a strong  acceleration of the particles
in the initial plasma. Such large values of the longitudinal
momentum of electrons (positrons) in the laboratory frame are due
to the transverse acceleration of  electrons (positrons)in the
moving frame. Performing the Lorentz transformation back to the
laboratory frame we obtain for the longitudinal momentum in the
laboratory frame of the initial electrons and positrons
$p_\|=\gamma_g[p_{\|o} +v_g(1+p_{\|o}^{ 2
}+|{\mathbf{a}}^{2}|)^{1/2}]\approx \gamma_g v_g |\mathbf{a}|$
where we used $|{\mathbf{a}}|\gg | p_{\|o}|$.

In summary,  the   production  of electron-positron pairs by the
electromagnetic wave propagation in the plasma  leads to the
up-shifting of the  wave frequency and  to the damping of the wave
amplitude and changes the polarization state of the wave. In order
to illustrate these effects analytically, we consider two limiting
cases of the electric field evolution, which can be clearly
distinguished  on the basis of the results shown in Fig. 1. First,
we will consider the case of a fast changing field when the
initial electric field amplitude is so large  that the most  of
the electron-positron pairs is produced instantly. This case can
be illustrated by the results shown in Fig. 1c. Then,  we will
consider the case of a slowly changing field, when the pair
production rate is relatively small and all the parameters of the
wave change slowly as illustrated in  Fig. 1a.

\section{Fast changing electromagnetic field}\label{L1}

We assume that the amplitude of the initial electric field is so
large that the electron-positron pair production and the
consequent change of the properties  of the wave electric field
occur almost instantaneously. In this case we may approximate the
time dependence of the source term as
\begin{equation}
q_{\alpha }(E,p,t)=\tau \delta (t)q_{\alpha }(E_{0},p),
\end{equation}
where $E_{0}$ is the initial amplitude of the electric field and
$\tau$ is a characteristic time defined so that the total number
$N$ of pairs produced is kept constant and given by $\tau
({|\mathbf{e}_{0}|^{2}}/{4\pi ^{3}})\exp{[ -{\pi
}/{|\mathbf{e}_{0}|}]}$ per Compton 3-volume $l_c^3$. Then, it is
possible to carry out the integration over $t^{\prime }$ in the
r.h.s. of Eq.(\ref{maxwell}). As a result we obtain
\begin{equation}
\frac{d^{2}\mathbf{a}}{dt^{2}}+\omega
_{p}^{2}\frac{\mathbf{a}(t)}{[ {1+p_{\Vert
0}^{2}+{a}^{2}(t)}]^{1/2}}=-\kappa \tau
\frac{|\mathbf{e}_{0}|^{2}}{8\pi ^{3}}\exp \left[ -\frac{\pi }
{|\mathbf{e}_{0}|}\right] \frac{\mathbf{a}(t)-\mathbf{a}_{0}
}{[{1+|\mathbf{a}(t)-\mathbf{a}_{0}|^{2}}]^{1/2}} \label{fast}
\end{equation}

We  note that $\mathbf{a}_{0}$ in Eq.(\ref{fast}) and the initial
value $\mathbf{a}(0)$ of the amplitude of the vector potential in
Eq.(\ref{fast})  are connected in an indirect way since the
initial conditions for Eq.(\ref{fast}) are determined by   the
value of the electric field {\it after} the main part of  the pair
production has already taken place when the amplitude of the field
has been reduced so that it is no longer capable of producing a
significant amount of pairs. Therefore  the values
$\mathbf{a}_{0}, \mathbf{e}_{0}$  that enter  the r.h.s. of
Eq.(\ref{fast}) should be considered as referring to  the vector
potential and to the  electric field at the instant when the pairs
are created.   From these values  we can deduce the value of
$\mathbf{a}(0)$  using  energy conservation and taking into
account the polarization current.

In order to understand the basic properties of the  electric field
evolution described by Eq.(\ref{fast}), it is convenient to
linearize it. To do so, we assume that $a\ll 1$ and obtain
\begin{equation}
\mathbf{a}^{\prime \prime }(t)+(\omega _{p}^{\prime 2}+\kappa
N)\mathbf{a} (t)-\kappa
N\mathbf{a}_{0}=0, \label{lin}
\end{equation}
where  $\omega _{p}^{\prime 2}=\omega _{p}^{2}/(1+p_{\Vert
0}^{2})^{1/2}$. We see that Eq.(\ref{lin}) describes oscillations
with  frequency  $(\omega_{p}^{\prime 2}+\kappa N)^{1/2}$. Taking
$\mathbf{a}_{0}=(a_{0z},0)$ we obtain
\begin{eqnarray}
a_{x}(t) &=&\frac{a_{0x}\kappa N}{\omega _{p}^{\prime 2}+\kappa
N}-\frac{a_{0x}\kappa N}{\omega
_{p}^{\prime 2}+\kappa N}\cos \left[ (\omega _{p}^{\prime 2}+\kappa
N)^{1/2}t\right] +a_{1}\cos \left[
(\omega _{p}^{\prime 2}+\kappa N)^{1/2}t\right] ,  \label{a_x} \\
a_{y}(t) &=&  -a_{1}\sin \left[
(\omega _{p}^{\prime 2}+\kappa N)^{1/2}t\right] ,  \label{a_y}
\end{eqnarray}
where  the initial conditions are $\mathbf{a}(0)=(a_{1},0)$,
$\mathbf{a}^{\prime }(0)=(0,(\omega _{p}^{\prime 2}+\kappa
N)^{1/2}a_{1})$. Thus,  Eqs. (\ref{a_x},\ref{a_y})  describe an
elliptically polarized  field oscillating  with an upshifted
frequency,  as consistent with the numerical solution of
Eq.(\ref{maxwell}).

\section{Slowly changing electromagnetic field}\label{L2}

We assume that the pair production rate is small, so  that the
parameters of the electromagnetic wave change  slowly. Then,  we
can neglect $a(t)$ with respect to $a(t^{\prime })$, since, as the
wave amplitude decreases with time,   $a(t)$ becomes increasingly
small compared to  $a(t^{\prime })$. Then, after differentiation
with respect to time, Eq.(\ref{maxwell}) reads
\begin{equation}\label{slow3}
\frac{d^{3}\mathbf{a}(t)}{dt^{3}}+\omega
_{p}^{2}\frac{d}{dt}\frac{\mathbf{a}(t)}{(1+p_{\Vert
0}^{2}+\mathbf{a}^{2}(t))^{1/2}}=\kappa
\frac{\mathbf{a}(t)}{[1+{a}^{2}(t)]^{1/2}}|\mathbf{e}(t)|^{2}\exp
\left[ -\frac{\pi
}{|\mathbf{e}(t)|}\right].
\end{equation}
If we linearize Eq. (\ref{slow3})  and set the pair production
rate equal to constant, we obtain
\begin{equation}\label{slow3l}
\mathbf{a}^{\prime \prime \prime }(t)+\frac{\omega
_{p}^{2}}{(1+p_{\Vert 0}^{2})^{1/2}}\mathbf{a}^{\prime }(t)-\kappa
   W \mathbf{a}(t)=0,
\end{equation}
where $W=|\bar{e}|^{2}\exp {( -{\pi}/{|\bar{e}|})}$ and $\bar{e}$
is the electric field amplitude averaged over the field evolution.
The equations for the  two components of the vector-potential are
decoupled and thus the polarization state is preserved. The
solution of Eq. (\ref{slow3}) is
\begin{equation}
a_{x,y}(t)=C_{1x,y}\exp (-i\omega _{1}t)+C_{2x,y}\exp (-i\omega
_{2}t)+C_{3x,y}\exp (-i\omega _{3}t),
\end{equation}
where the frequencies $\omega _{i}$ ($i=1,2,3$) are the roots of
the third order polynomial equation $y^{3}- y\omega
_{p}^{2}/(1+p_{\Vert 0}^{2})^{1/2} +i \kappa W =0$ and the
constants $C_{ix,y}$ are determined by the initial conditions. It
is obvious from the form of the polynomial equation that one of
its roots is imaginary  and positive, while the other two are
complex. The two terms in the solution with complex frequencies
describe a damped wave, while the third term corresponds to  a
spurious,  exponentially growing, term that can be  excluded by an
appropriate choice of the initial conditions.

\section{Conclusion and discussions}\label{CC}

In the present paper we considered the problem of the backreaction
of the produced electron-positron pairs on the electromagnetic
wave. We showed that there is a  loss of wave energy due to the
pair production and  acceleration of these pairs in the
electromagnetic field of the wave.

We studied the propagation  of a relativistically strong
electromagnetic wave  in an  underdense electron-positron plasma.
As  is well known, a plane wave does not produce electron-positron
pairs in vacuum. However  the situation changes in a plasma. Due
to the fact that in plasma the first field invariant $\mathcal{F}$
does not vanish, a plane wave can produce pairs that backreact  on
the wave. In order to describe the behaviour of the  electrons and
positrons, we adopted a kinetic plasma description and used  the
relativistic Boltzmann-Vlasov equation. Solving  this equation we
obtained  the distribution functions of electrons and positrons
which we used to derive the current density that enters the r.h.s.
of Maxwell equation for the electric field. We  note that all the
calculations were carried out in the reference frame moving with
the group velocity of the wave. In this frame there is no magnetic
field as the wave has only an  electric field.

The solutions of the Maxwell equation for the evolution of the
electric field in the plasma for different initial amplitudes are
shown in Fig. 1. We see that, when the process of
electron-positron pair production is taken into account, the
evolution of the vector-potential in the plasma leads to the
damping of the wave, accompanied by a nonlinear shift of its
frequency. The damping of the vector-potential is due to the pair
production which  takes away a portion of the wave energy. This
process is followed by the increase of plasma density which leads
to the up-shifting of the  wave  frequency.

The damping of the vector-potential amplitude, together with its
frequency increase, gives rise to an interesting phenomenon -- the
decrease of the wave amplitude followed by the increase of its
frequency. Pair production  decreases the wave amplitude while the
growth of plasma density all over the space causes the increase of
the wave frequency which is proportional to square root of the
density. This effect resembles the case  when the electromagnetic
wave propagates in an ionizing medium \cite{ionizationth}.

Since the pair production rate depends on the field amplitude
exponentially, an unbalanced damping of the field components
occurs and leads to the change of the wave polarization (see Fig.
2). One can clearly see the change of the electric field
polarization from circular to elliptic  and the decrease  of one
of the two electric field components.

Finally,  we considered two limiting cases in order to identify
the properties and the exact causes of this nonlinear behaviour of
the wave. In the first case the initial amplitude of the electric
field is so large  that the pair production and the change of the
wave properties occur instantaneously. In the second case the pair
production rate is small so that the properties of the wave change
slowly. In first case we obtain an electric field oscillating with
a new frequency and amplitude. The frequency  increase is due to
the fact that the density of plasma increases because of the pair
production. In the second case we obtain  a damped wave with an
amplitude  that decreases slowly  with time. We see that the
process of electron-positron pair production leads to the damping
of the wave. There is a nonlinear shift of its frequency due to
the decrease of its amplitude: as a result the frequency is
up-shifted.

\acknowledgments

The authors would like to acknowledge fruitful discussions with N.
B. Narozhny, V. D. Mur, L. B. Okun and V. S. Popov. This work was
supported in part by the Russian Fund for Fundamental Research
under projects 03-02-17348, the Federal Program of the Russian
Ministry of Industry, Science and Technology grant No
40.052.1.1.1112.

\section*{Appendix: Properties of the polarization current}

The form of the polarization current can be derived from the
energy-conservation law. In order to calculate the expression for
the polarization current we calculate the second moment of the
kinetic equation (\ref{kinetic}).
\begin{equation}
\sum\limits_{\alpha =\pm }\int \left[ (m^{2}+p^{2})^{1/2}-m\right]
\left\{ \frac{\partial f_{\alpha }}{\partial t}+e_{\alpha
}\mathbf{E}\frac{\partial f_{\alpha }}{\partial
\mathbf{p}}\right\} d^{3}p=\sum\limits_{\alpha =\pm }\int \left[
(m^{2}+p^{2})^{1/2}-m\right] q_{\alpha
}(\mathbf{E},\mathbf{p})d^{3}p,
\end{equation}
which we write as
\begin{equation}
\frac{\partial K}{\partial t}-\mathbf{j}_{cond}\mathbf{E}=\Sigma
-2 m\frac{\partial n}{\partial t}, \label{2moment}
\end{equation}
  where
\[
K=\sum\limits_{\alpha =\pm }\int \left[ (m^{2}+p^{2})^{1/2}-m\right]
f_{\alpha }(\mathbf{p},t)d^{3}p
\]
and
\[
j_{cond}=e\int
\frac{\mathbf{p}}{(m^{2}+\mathbf{p}^{2})^{1/2}}\sum\limits_{\alpha
=\pm }e_{\alpha
}f_{\alpha }(\mathbf{p},t)d^{3}p
\]
is the conduction current. The second  moment of the r.h.s. of
kinetic equation r.h.s. gives two terms. One of them, $-2m\,
\partial n/\partial t$, is related  to the rest energy increase
due to the pair production, and the another is equal to
\[
\Sigma =\sum\limits_{\alpha =\pm }\int (m^{2}+p^{2})^{1/2}q_{\alpha
}(\mathbf{E},\mathbf{p})d^{3}p.
\]
We can thus  rewrite Eq. (\ref{2moment}) in the following form
\begin{equation}
\frac{\partial }{\partial t}\left( K+2 n m \right)
=\mathbf{j}_{cond}\mathbf{E} +\Sigma ,  \label{energy p}
\end{equation}
where the expression inside the brackets correspond to the full
energy of the  plasma particles. The energy balance equation
for  the electromagnetic field in the moving reference, frame
where only the electric field is present, can be obtained by
multiplying Eq. (\ref{max}) by the vector $\mathbf{E}$
\begin{equation}
\frac{\partial }{\partial t}\left( \frac{\mathbf{E}^{2}}{8\pi }\right)
=-\mathbf{j}_{cond}\mathbf{E}-\mathbf{j}_{pol}\mathbf{E}.  \label{energy f}
\end{equation}
  Here we took into account that  no external current is
present in our case. Adding  Eqs. (\ref{energy p}) and
(\ref{energy f}) we obtain
\begin{equation}\label{eneT}
\frac{\partial }{\partial t}\left( K+2
n+\frac{\mathbf{E}^{2}}{8\pi }\right) =-
\mathbf{j}_{pol}\mathbf{E}+\Sigma .
\end{equation}
The expression inside the brackets in Eq.(\ref{eneT})  is the
energy of the system of the electromagnetic field and the plasma
electrons and positrons. Energy conservation requires the r.h.s.
of Eq.(\ref{eneT}) to be  zero. Thus the  polarization current
should be of the form
\begin{equation}
\mathbf{j}_{pol}=\frac{\mathbf{E}}{|\mathbf{E}|^{2}}\sum\limits_{\alpha =\pm
}\int (m^{2}+\mathbf{p}^{2})^{1/2}q_{\alpha }(\mathbf{p},t)d^{3}p.
\end{equation}


\begin{thebibliography}{99}
\bibitem{Saut}  F. Sauter, Z. Phys. \textbf{98}, 714 (1931), \textbf{98},
714 (1931).

\bibitem{el-pos}  W. Heisenberg and H. Z. Euler, Z. Phys. \textbf{98}, 714
(1936).

\bibitem{Schwinger}  J. Schwinger, Phys. Rev. \textbf{82}, 664 (1951).

\bibitem{B-I}  E. Brezin and C. Itzykson, Phys. Rev. D \textbf{2}, 1191
(1970).

\bibitem{Popov71}  V. S. Popov, JETP Lett. \textbf{13}, 185 (1971); Sov.
Phys. JETP \textbf{34}, 709 (1972).

\bibitem{Popov73}  V. S. Popov, JETP Lett. \textbf{18}, 255 (1973); Sov. J.
Nucl. Phys. \textbf{19}, 584 (1974).

\bibitem{PopovMarinov}  M. S. Marinov and V. S. Popov, Sov. J. Nucl. Phys.
\textbf{16}, 449 (1973).

\bibitem{NN}  N. B. Narozhny and A. I. Nikishov, Sov. Phys. JETP
\textbf{38}, 427 (1974).

\bibitem{Bunkin}  F. V. Bunkin and I. I. Tugov, Dokl. Akad. Nauk SSSR
\textbf{187}, 541 (1969).

\bibitem{Troup}  G. J. Troup and H. S. Perlman, Phys. Rev. D \textbf{6},
2299 (1972).

\bibitem{Mourou}  T. Tajima, G. Mourou, Phys. Rev. ST-AB \textbf{5}, 031301
(2002); arXiv physics/0111091.

\bibitem{Burke}  D.L. Burke, \textit{et al.}, Phys. Rev. Lett. \textbf{79},
1626 (1997).

\bibitem{LightIntensification}  S. V. Bulanov, T. Zh. Esirkepov, and T.
Tajima, Phys. Rev. Lett. \textbf{91}, 085001 (2003).

\bibitem{ssbul}  S. S. Bulanov, Phys. Rev. E. \textbf{69}, 036408 (2004).

\bibitem{BNMP}  S. S. Bulanov, N.B. Narozhny, V.D. Mur, V.S. Popov,
hep-ph/0403163.

\bibitem{Matsui}  K. Kajantie and T. Matsui, Phys. Lett. \textbf{164B}, 373
(1985).

\bibitem{Matsui1}  G. Gatoff, A. K. Kerman, and T. Matsui, Phys. Rev. D
\textbf{36}, 114 (1987).

\bibitem{Kluger}  Y. Kluger, J. M. Eisenberg, and B. Svetitsky, Phys. Rev.
Lett. \textbf{67}, 2427 (1991); Y. Kluger, J. M. Eisenberg, and B.
Svetitsky, Phys. Rev. D \textbf{45}, 4659 (1992); Y. Kluger, J. M.
Eisenberg, and B. Svetitsky, Int. J. Mod. Phys. E \textbf{2}, 333 (1993); Y.
Kluger, E. Mottola, and J. M. Eisenberg, Phys. Rev. D \textbf{58}, 125015
(1998).

\bibitem{Schmidt}  S. Schmidt, D. Blanschke, G. Ropke, S. A. Smolyansky, A.
V. Prosorkevich, and V. D. Toneev, Int. J. Mod. Phys. E \textbf{7}, 709
(1998); S. Schmidt, D. Blanschke, G. Ropke, S. A. Smolyansky, A. V.
Prosorkevich, and V. D. Toneev, Phys. Rev. D \textbf{59}, 094005 (1999); J.
C. R. Bloch, V. A. Myserny, A. V. Prosorkevich, C. D. Roberts, S. M.
Schmidt, S. A. Smolyansky, and V. D. Toneev, Phys. Rev. D \textbf{60},
116011 (1999).

\bibitem{Dolby}  C. E. Dolby and S. F. Gull, Annals Phys. \textbf{297}, 315
(2002).

\bibitem{Ruffini}  R. Ruffini, L. Vitagliano, and S. S. Xue, Phys. Lett. B
\textbf{559}, 12 (2003).

\bibitem{avetissian}  H. K. Avetissian, A. K. Avetissian, G. F. Mkrtchian,
and Kh. V. Sedrakian, Phys. Rev. E \textbf{66}, 016502 (2002).

\bibitem{AP}  A.~I.~Akhiezer, R.~V.~Polovin, Sov Phys. JETP \textbf{30}, 915
(1956).

\bibitem{far}
I.V. Smetanin, D. Farina, J. Koga, K. Nakajima, S.V. Bulanov, Phys.
Lett. {\bf A 320},438
(2004).


\bibitem{Popov}  V. S. Popov, JETP Lett. \textbf{13}, 185 (1971); Sov. Phys.
JETP \textbf{34}, 709 (1972); V. S. Popov, JETP Lett. \textbf{18}, 255
(1973); Sov. J. Nucl. Phys. \textbf{19}, 584 (1974).

\bibitem{ionizationth} S. C. Wilks, J. M. Dawson, and W. B. Mori, Phys.
Rev. Lett. \textbf{61}, 337 (1988);
A. V. Kim, S. F. Lirin, A. M. Sergeev, E. V. Vanin, and L.
Stenflo, Phys. Rev. A \textbf{42}, 2493 (1990); E. Esarey, G. Joyce, and P.
Spranglef, Phys. Rev. A \textbf{44}, 3908 (1991);
V. B. Gil'denburg, V. I.
Pozdnyakova, and I. A. Shereshevskii, Phys. Lett. A \textbf{203}, 214
(1995); E. Conejero Jarque, F.  {Cornolti},and A. {Macchi},
J. Phys. B, \textbf{33}, 1 (2000).

\bibitem{ionizationexp}
S. P. Kuo, Phys. Rev. Lett. \textbf{65}, 1000 (1990);
N. Yugami, T. Niiyama, T. Higashiguchi, H. Gao, S. Sasaki, H. Ito,
and Y. Nishida, Phys. Rev. E \textbf{65}, 036505 (2002).
\end{thebibliography}
\end{document}